\documentclass[10pt,twocolumn,journal]{IEEEtran}
\usepackage{setspace}
\usepackage{amsmath}
\usepackage{algorithmic}
\usepackage{graphicx}
\usepackage{epstopdf}
\usepackage{cite}
\usepackage{algorithm}
\usepackage{amssymb}
\usepackage{epsfig}
\usepackage{array}
\usepackage{subfigure}
\usepackage{bm}
\usepackage{color}
\usepackage{latexsym}
\usepackage{multirow}

\def\bfC{{\mathbf{c}}}

\begin{document}
%
\title{Object Shape Approximation \& Contour Adaptive Depth Image Coding for Virtual View Synthesis}

\author{Yuan Yuan~\IEEEmembership{Student Member,~IEEE},
Gene Cheung~\IEEEmembership{Senior Member,~IEEE},
Patrick Le Callet~\IEEEmembership{Senior Member,~IEEE},
Pascal Frossard~\IEEEmembership{Senior Member,~IEEE},
H. Vicky Zhao~\IEEEmembership{Senior Member,~IEEE}

\begin{small}

\thanks{Y. Yuan is with the Electrical and Computer Engineering Department, University of Alberta, T6G 1H9, Edmonton, AB, Canada (e-mail: yyuan13@ualberta.ca).}

\thanks{G. Cheung is with the National Institute of Informatics, Graduate University for Advanced Studies, Tokyo 101-8430, Japan (e-mail: cheung@nii.ac.jp).}

\thanks{P. Le Callet is with Polytech Nantes/Universite de Nantes, IRCCyN/IVC. rue Christian Pauc La Chantretir, BP 50609 44306 Nantes Cedex 3 (e-mail: patrick.lecallet@univ-nantes.fr).}

\thanks{P. Frossard is with Signal Processing Laboratory (LTS4), Ecole Polytechnique F\'{e}d\'{e}rale de Lausanne (EPFL), CH-1015 Lausanne, Switzerland (e-mail: pascal.frossard@epfl.ch).}

\thanks{V. Zhao is with Department of Automation, Tsinghua University, State Key Lab of Intelligent Technologies and Systems, Tsinghua National Laboratory for Information and Science and Technology (TNList), Beijing, P.R.China (e-mail: vzhao@tsinghua.edu.cn).}

\end{small}
}

\maketitle

\begin{abstract}
A depth image provides partial geometric information of a 3D scene, namely the shapes of physical objects as observed from a particular viewpoint. 
This information is important when synthesizing images of different virtual camera viewpoints via depth-image-based rendering (DIBR). 
It has been shown that depth images can be efficiently coded using contour-adaptive codecs that preserve edge sharpness, resulting in visually pleasing DIBR-synthesized images. 
However, contours are typically losslessly coded as side information (SI), which is expensive if the object shapes are complex.

In this paper, we pursue a new paradigm in depth image coding for color-plus-depth representation of a 3D scene: we pro-actively simplify object shapes in a depth and color image pair to reduce depth coding cost, at a penalty of a slight increase in synthesized view distortion.
Specifically, we first mathematically derive a distortion upper-bound proxy for 3DSwIM---a quality metric tailored for DIBR-synthesized images.
This proxy reduces inter-dependency among pixel rows in a block to ease optimization.
We then approximate object contours via a dynamic programming (DP) algorithm to optimally trade off coding cost of contours using arithmetic edge coding (AEC) with our proposed view synthesis distortion proxy. 
We modify the depth and color images according to the approximated object contours in an inter-view consistent manner. These are then coded respectively using a contour-adaptive image codec based on graph Fourier transform (GFT) for edge preservation and HEVC intra.
Experimental results show that by maintaining sharp but simplified object contours during contour-adaptive coding, for the same visual quality of DIBR-synthesized virtual views, our proposal can reduce depth image coding rate by up to $22\%$ compared to alternative coding strategies such as HEVC intra.
\end{abstract}



%
\IEEEpeerreviewmaketitle

\section{Introduction}
\label{sec:intro}
The advent of depth sensing technologies like Microsoft Kinect has eased the acquisition of depth images (namely per-pixel distances between physical objects in a 3D scene and a capturing camera). 
Each depth map provides important geometric information---object shapes and contours as observed from a camera viewpoint---which is used, together with a color image from the same viewpoint, to synthesize virtual viewpoint images via \textit{depth-image-based rendering} (DIBR)~\cite{tian09}. 
This ability of virtual view synthesis enables a plethora of 3D imaging applications, including free viewpoint TV~\cite{tanimoto11}, or immersive video conferencing~\cite{zhang09}.

\begin{figure}[t]
\centering
\centerline{\includegraphics[width=8.6cm]{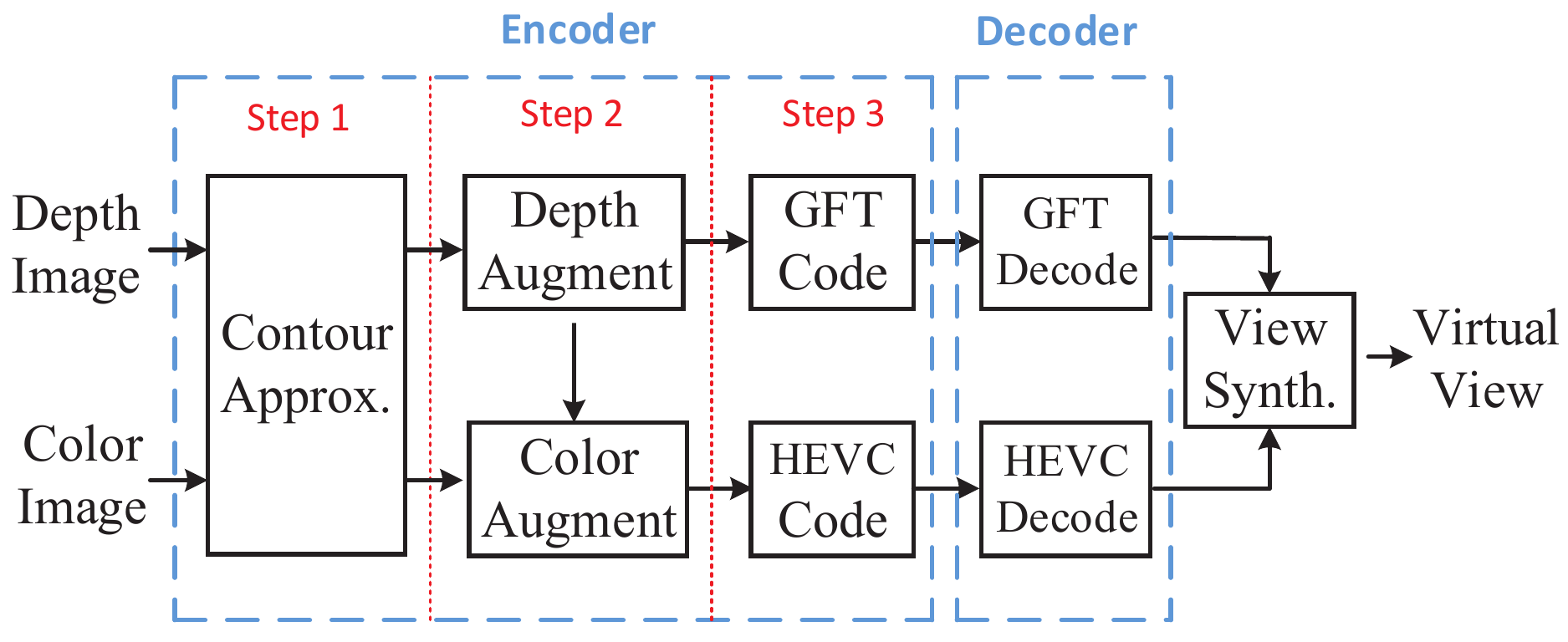}}
\caption{Overview of our proposed color-plus-depth image coding system.}
\label{fig:sysReview}
\end{figure} 

To enable decoder-side virtual view synthesis, depth and color image pairs from the same viewpoints must be compressed for network transmission.
Traditional image codecs like JPEG, H.264 and HEVC~\cite{sullivan2012overview} employ fixed block transforms like DCT, where coarser quantization of transform coefficients at low bit rates will result in blurring of sharp edges.  
It has been demonstrated that blurring of object contours in a depth map leads to unpleasant bleeding artefacts in images synthesized via DIBR \cite{kim10}. 
Thus state-of-the-art depth image coding algorithms employ contour-adaptive transforms \cite{shen10pcs,weihu12icip,hu2015tip,chao15} and wavelets \cite{maitre08} to preserve sharp object contours, which are losslessly coded as side information (SI) separately. 
However, the SI coding cost can be expensive at low rates, amounting to $50\%$ of the total depth bit budget in some cases \cite{hu2015tip}. 

In response, in this paper we pursue a new paradigm in depth image coding for color-plus-depth representation of a 3D scene: 
\textit{we pro-actively simplify complex object shapes in a depth and color image pair to lower depth coding cost, at a controlled increase in synthesized view distortion.}
This means that as the bit budget becomes stringent, actual shapes of physical objects in the scene are simplified, but rendering of the objects remains sharp and natural for human observation. 

Specifically, we execute our proposed object shape approximation in our color-plus-depth coding system as follows. 
(See Fig.\;\ref{fig:sysReview} for an overview.)
Given an input color-plus-depth image pair from the same viewpoint, we first approximate object contours via a dynamic programming (DP) algorithm to optimally trade off the cost of contours coded using arithmetic edge coding (AEC)~\cite{daribo2012icip} and the synthesized view distortion induced due to contour approximation. 
The color and depth images are then modified according to the approximated object contours to ensure inter-view consistency.
Finally, the modified depth image is coded using a contour-adaptive image codec based on graph Fourier transform (GFT) for edge preservation~\cite{hu2015tip}, and the modified color image is coded by HEVC intra. 
At the decoder, the decoded depth and color images are then used for free-view synthesis via DIBR.

To measure the induced synthesized view distortion, we propose to use 3DSwIM~\cite{battisti2015}---a quality metric tailored specifically for DIBR-synthesized images. 
However, the complex definition of 3DSwIM makes it too expensive to directly quantify distortion due to contour approximation. 
In response, we mathematically derive a local distortion proxy that serves as an upper bound of 3DSwIM with reduced inter-dependencies across pixel rows to ease optimization.

Through extensive experiments, we demonstrate that by maintaining sharp but simplified object contours, our proposed coding scheme requires less coding bits to achieve the same visual quality of DIBR-synthesized virtual views. Specifically, our scheme can reduce depth coding rate by up to $18\%$ compared to unmodified depth map coding by \cite{hu2015tip}, and up to $22\%$ compared to HEVC intra. 
Subjective quality evaluation also shows that by maintaining sharp edges in the decoded depth images, rendered visual quality is also more visually pleasing.

The outline of the paper is as follows. 
We first overview related work in Section~\ref{sec:related}. 
We then discuss contour coding rate and distortion in Section~\ref{sec:rate} and Section~\ref{sec:dist}, respectively. Based on them, a rate-distortion (RD) optimal method for depth contour approximation is proposed in Section~\ref{sec:app}. 
Image coding using the approximated object contours will be discussed in Section~\ref{sec:coding}.
Finally, experiments and concluding remarks are presented in Section~\ref{sec:results} and \ref{sec:conclude}, respectively.

\section{Related Work}
\label{sec:related}
We divide our discussion of related work into three sections. We first discuss existing literature on depth image coding. 
We then overview previous work on image contour coding. 
Finally, we discuss existing work on shape approximation.

\subsection{Depth Image Coding}

Typical image coding algorithms employing fixed block-based transforms like DCT~\cite{schwarz2007overview}~\cite{skodras2001jpeg} can only compactly represent image patches with horizontal and vertical contours well. Directional transforms \cite{zeng200icme} can adapt to diagonal contours, but cannot deal with more arbitrarily shaped contours such as ``L" and ``V". 
Practically, that means coarse quantization of the high-frequency components in these transforms at low bit rates will lead to blurring of arbitrarily shaped contours in the reconstructed depth image.

Observing that depth image contours play an important role in the quality of the DIBR-synthesized view~\cite{kim10}, contour-preserving image coding algorithms have been proposed. 
Much of these work exploits depth images' \textit{piecewise smooth} (PWS) characteristics: smooth interior surfaces separated by sharp contours.
\cite{morvan06} models depth images with piecewise linear functions (platelet) in each block. 
However, the representation inherently has a non-zero approximation error even at high rates, since depth images are not strictly piecewise linear but PWS. 
\cite{sanchez09} and~\cite{maitre08} propose contour-adaptive wavelets, and
\cite{shen10pcs} proposes block-based unweighted GFT for depth map coding. 
Extending \cite{shen10pcs}, \cite{hu2015tip} searches for an optimal weighted graph for GFT-based image coding in a multi-resolution framework. 
In all these works, detected edges are encoded losslessly as SI, which can cost up to 50\% of the total budget at low rates. 

Since depth maps are usually used for synthesizing virtual views at the decoder side and are not directly viewed, it is necessary to consider the resulting synthesized view distortion, instead of depth distortion itself, for RD optimization during depth map coding~\cite{kim10,kim09}.
How depth distortion related to synthesized view distortion has been investigated \cite{oh11jst,zhang2013regional,yuan2014novel}. 
For example, \cite{oh11jst} proposes a synthesized view distortion function as the multiplication of depth distortion and local color information.
However, these methods assume PSNR is the quality metric for the synthesized view when compressing depth images, which has been demonstrated to not correlate well with actual human perception \cite{battisti2015}.

In \cite{battisti2015}, the authors show that while objects in a synthesized view may be slightly shifted due to the DIBR-rendering process, the overall visual quality of the image may still be acceptable. 
However, this type of artifacts is penalized by pixel-by-pixel based quality metric such as PSNR. 
They thus propose a 3D synthesized view image quality metric (3DSwIM) tailored specifically for artifacts in DIBR synthesized views. 
They conclude that this metric has a higher correlation with subjective quality assessment scores than PSNR.
Using 3DSwIM as the chosen metric, in this work we approximate object contours and augment depth / color image pairs for more efficient coding of 3D image content.

There are other pre-processing methods that attempt to improve video / image coding efficiency. 
The work in~\cite{algazi1995preprocessing,kim1998block,shen1999video,li2006adaptive} apply filters on input videos / images to remove noise.
\cite{doutre2009color} corrects the color value of multiview videos to make them consistent, and \cite{van2015improving} evaluates the coding performance of applying a texture image guided filter on the estimated low quality depth map. 
Different from these works, we propose to modify the geometry structure of a 3D scene to improve color-plus-depth image coding efficiency.

Besides the well known and accepted color-plus-depth image representation of the 3D scene, recently a graph-based representation (GBR) for 3D scene is proposed \cite{maugey2015graph}. 
In GBR, the disparity information between two viewpoint images is compactly described as a graph; a complex shaped object would result in a complicated graph in GBR.
Thus, though in this paper we show only that contour approximation can lead to coding gain in color-plus-depth image representation of a 3D scene, in theory similar contour approximation would also lead to coding gain in GBR representation of the same scene.

\subsection{Object Contour Coding}

Freeman chain code~\cite{Freeman61} is widely used to efficiently encode object contours~\cite{akimov2007lossless,liu2005efficient,zhou2006new}. 
Usually, object contours are first convert into a sequence of symbols, where each symbol is from a finite set with four or eight possible absolute directions. 
Alternatively, the relative directions between two neighbouring directions, which is also known as \textit{differential chain code} (DCC)~\cite{freeman1978application} can also be used. 
Given the probability of each symbol, the contour chain code are entropy-coded losslessly using either Huffman~\cite{liu2005efficient} or arithmetic coding~\cite{witten1987arithmetic}.


The works in~\cite{daribo2012icip}~\cite{daribo2014tip} introduced an \textit{arithmetic edge coding} (AEC) method employing a linear \textit{geometric model} to estimate the probability of next edge. 
Given a small window of previous edges, they first construct a best-fitting line that minimizes the sum of squared distances to the end point of each edge. Then the probability for the next edge direction is assigned based on the angle difference between the edge direction and the best fitted line,
which is subsequently used as context for arithmetic coding. 
The assigned probability for current edge only depends on previous encoded edges. In our work, we improve the AEC model by adding a distance term to more appropriately assign probabilities for contour coding. 
Details will be discussed in section \ref{sec:rate}.

\subsection{Image Contour Approximation}

There exist some contour approximation methods in the compression literature.
The work in~\cite{katsaggelos1998mpeg} proposes a polygon / spline-based representation to approximate a contour. 
Vertices are encoded differentially and polygon curves between neighbouring vertices are considered to approximate the original contours.
 Later, the authors in~\cite{yun2001vertex,kuo2007new,lai2010arbitrary,lai2014operational} improve this vertex-based shape coding method.
All of these works take the maximum or mean distance between original and approximated contours as distortion measure, which is not appropriate for assessment of visual quality of synthesized views. 

Some other works approximate contours from the chain code representation.
The method in~\cite{yeh2002scalable} saves coding bits by omitting two neighboring turning points if the slope between them is nearly vertical or horizontal.
The algorithm~\cite{daribo2014tip} and \cite{zahir2007new} replace some pre-defined irregular edge patterns with a smooth straight line. 
In our previous work \cite{yuan2015contour}, the edge direction with the largest estimated probability computed by AEC~\cite{daribo2012icip} is selected to reduce coding cost.
None of these works consider synthesized view distortion induced by contour approximation. 
In this work, we approximate contours using the 3DSwIM metric to control the induced synthesized view distortion, which leads to more pleasant results in view synthesis.





\section{Contour Coding Rate: Arithmetic Edge Coding}
\label{sec:rate}

It is observed that depth maps exhibit the PWS characteristic \cite{shen10pcs,weihu12icip}, \textit{i.e.}, smooth surfaces divided by sharp edges. Given that the smooth portions can be efficiently coded using edge-adaptive wavelets \cite{maitre08} and transforms \cite{shen10pcs,hu2015tip}, we are interested here only in the rate to encode the set of boundaries (object contours) separating foreground objects from the background. Hence, before we proceed with the contour approximation problem, we first review one previous contour coding scheme---\textit{arithmetic edge coding} (AEC).

\begin{figure}[htb]
\begin{minipage}[b]{.23\textwidth}
  \centering
  \centerline{\includegraphics[width=4.0cm]{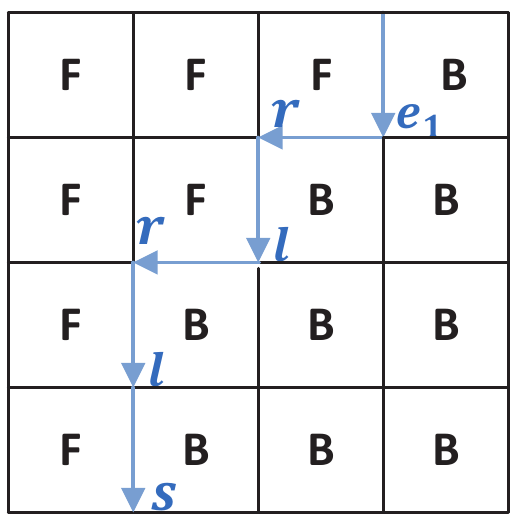}}
  \centerline{(a) Between-pixel edges}\medskip
\end{minipage}
\hfill
\begin{minipage}[b]{.23\textwidth}
  \centering
  \centerline{\includegraphics[width=3.8cm]{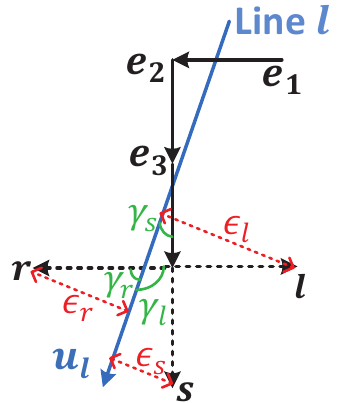}}
  \centerline{(b) Edge prediction}\medskip
\end{minipage}
\caption{(a) Edges in a $4\times4$ block that separate foreground (F) from background (B). (b) Given edges $\{e_3,e_2,e_1\}$ to estimate the probability of edge $e_4$. With the best-fitting line $l$ and its direction $u_l$, the angel difference $\gamma_\theta$ and distance $\epsilon_\theta$ for $e_4 = \theta, \theta \in \mathcal{A}$ is illustrated.  }
\label{fig:predict}
\end{figure}

Given a depth image, we first detect edges via a gradient-based method \cite{hu2015tip}, where the edges exist \textit{between} pixels and outline contours of physical objects. 
Fig.\;\ref{fig:predict}(a) shows an example of a $4 \times 4$ block with \textit{between-pixel edges}---each edge exists between a pair of vertically or horizontally adjacent pixels---separating foreground and background depth values. 
The set of contiguous edges composing a contour is described by a sequence of symbols $\mathcal{E} = \{e_T, \ldots, e_1\}$ of some length $T$. 
Each $e_t$, $t > 1$, is chosen from a size-three alphabet---$e_t \in \mathcal{A} = \{ \mathtt{l}, \mathtt{s}, \mathtt{r}\}$---representing \textit{relative} directions left, straight and right respectively, with respect to the previous edge $e_{t-1}$. 
In contrast, the first edge $e_1$ is chosen from a size-four alphabet---$e_1 \in \mathcal{A}^o = \{ \rightarrow, \downarrow, \leftarrow, \uparrow  \}$---denoting \textit{absolute} directions east, south, west and north respectively.
This contour representation is also known as \textit{differential chain code} (DCC)~\cite{Freeman61}. 

To efficiently encode a symbol $e_{t+1}$ in sequence $\mathcal{E}$, using a small window of $K$ previous edges $\{e_t,\ldots, e_{t-K+1}\}$ as context, AEC \cite{daribo2012icip,daribo2014tip} uses a linear regression model to estimate  probabilities $Pr(e_{t+1} = \theta)$ of the three possible directions $\theta \in \mathcal{A}$, which are subsequently used for arithmetic coding~\cite{witten1987arithmetic} of $e_{t+1}$. 
Specifically, using $\{e_t, \ldots, e_{t-K+1}\}$ they construct a \textit{best-fitting} line $l$ with direction $u_l$ via linear regression. 
By best-fitting, they mean that the line $l$ minimizes the sum of squared distances between the end point of each edge in $\{e_t, \ldots, e_{t-K+1}\}$ and $l$. 
See Fig.\;\ref{fig:predict}(b) for an illustration, where the line $l$ is the best fitting line given three edges $\{e_3,e_2,e_1\}$.

In \cite{daribo2012icip,daribo2014tip}, the angle $\gamma_\theta \in [0, \pi]$ between a relative direction $\theta \in \mathcal{A}$ and $u_l$ is first computed, then the probability $P(e_{t+1} = \theta)$ for edge $e_{t+1}$ is defined such that smaller $\gamma_{\theta}$ leads to larger $P(e_{t+1} = \theta)$. 
In our work, we consider also the minimum distance $\epsilon_\theta$ between the end point of $e_{t+1}$ and $l$ when determining $P(e_{t+1} = \theta)$.
Specifically, We define $P(e_{t+1} = \theta)$ as:
\begin{eqnarray}
\label{eq:prob}
P(e_{t+1} = \theta) = \frac{1}{2\pi I_0(\kappa)} \cdot \exp \left\{\kappa \cos \gamma_\theta\right\} \cdot \exp \left\{ -\frac{\epsilon_\theta^2}{2\omega^2} \right\}
\end{eqnarray}
where $I_0(\cdot)$ is the modified Bessel function of order 0. The parameter $1/\kappa$ is the variance in the circular normal distribution. 
$\omega$ is a chosen parameter to adjust the relative contribution of the distance term $\exp\{-\frac{\epsilon_\theta^2}{2\omega^2}\}$. 
The distance term is added to the Von Mises probability distribution model\footnote{http://en.wikipedia.org/wiki/Von\_Mises\_distribution} to differentiate the case where there exist two directions with the same $\gamma_\theta$ (\textit{e.g.}, a diagonal straight line). 
The computed probabilities $P(e_{t+1})$, synchronously computed at both the encoder and the decoder, are then used for arithmetic coding of the actual edge $e_{t+1}$. 

In words, (\ref{eq:prob}) states that the direction probability increases as the angle $\gamma_\theta$ and distance $\epsilon_\theta$ decrease.
Recall the previous example in Fig.\;\ref{fig:predict}(b),
with the best-fitting line $l$ and line direction $u_l$, the angle difference $\gamma_\theta$ and distance $\epsilon_\theta$ for $\theta \in \mathcal{A}$ are illustrated. 
The relative direction $\theta$ with the smallest angle and distance with respect to $u_l$ is $\mathtt{s}$ in this example. 

\section{Distortion: A Proxy of 3DSwIM}
\label{sec:dist}
\begin{figure*}[htb]
\centering
\centerline{\includegraphics[width=16.cm]{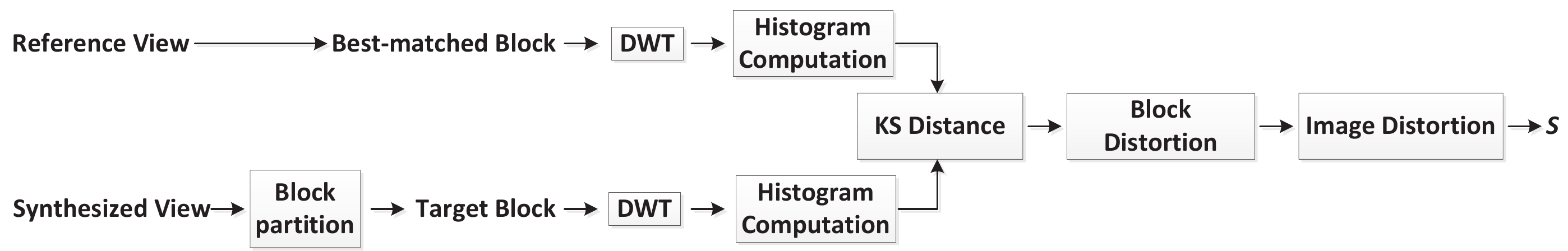}}
\caption{Block diagram of 3DSwIM, from~\cite{battisti2015}.}
\label{fig:3dswIM}
\end{figure*}

We next consider how to quantify distortion due to contour approximation.
We first overview \textit{3D Synthesized View Image Quality Metric} (3DSwIM)~\cite{battisti2015}, a new metric designed specifically to measure visual quality distortion of virtual view images synthesized via DIBR~\cite{tian09}.
Unfortunately, the 3DSwIM metric is difficult to optimize directly. 
We thus propose a simpler proxy that serves as an upper bound of 3DSwIM.

\subsection{3D Synthesized View Image Quality Metric}

The operations to compute 3DSwIM between a reference image and a DIBR-synthesized image is shown in the flow diagram in Fig.\;\ref{fig:3dswIM}. 
First, the synthesized image is divided into $B$ non-overlapping blocks, each of size $N \times N$. 
For a target block in the synthesized image, they search for the best-matched block (using mean square error as the matching criteria) in the reference image within a search window of size $2W$ pixels in the horizontal direction, via a search algorithm like \cite{dufaux1995motion}. 
This horizon search procedure is based on the observation that objects may be horizontally shifted during DIBR's 3D warping of pixels in rectified images.

According to \cite{solh20113vqm}'s analysis of DIBR distortions, the DIBR rendering errors are typically located near vertical edges of objects, and they appear as extraneous  horizontal details. 
Hence in~\cite{battisti2015}, the authors applied wavelet transform on each image block and measured the synthesized image degradation by analyzing the wavelet coefficients that describe the horizontal details. 
In the case of 3DSwIM, a multi-resolution 1D Haar wavelet transform (DWT)~\cite{burrus1997introduction} is applied on each row of a target block to extract the horizontal details. 

Since 3DSwIM is only concerned with horizontal image details, here we analyze the AC coefficients (details) of 1D DWT to compute a synthesized image quality measure. 
Performing a Haar DWT to each row of a $N \times N$ target block and its corresponding best-matched block in the reference image, each row generates an AC coefficient vector with length $(N-1)$. 
Stacking the coefficient vectors for different rows into a matrix, we construct two AC coefficient matrices $\bfC_s$ and $\bfC_o$ with size $N \times (N-1)$ corresponding to synthesized and reference blocks respectively.  

We then construct the histograms $\mathbf{H}_s$ and $\mathbf{H}_o$ for each AC coefficient matrices: we divide the coefficients range into $L$ bins of size $\tau$ and count how many coefficients fall into each bin, where $\tau$ is:
\begin{eqnarray}
\label{eq:binSize}
\tau  =  \frac{c_{max} - c_{min}}{L}, 
\end{eqnarray}
$c_{max} = \max \{\bfC_o, \bfC_s\}$ and $c_{min} = \min \{ \bfC_o, \bfC_s \}$ are the maximum and minimum values of the two AC coefficient matrices.

Finally, with the cumulative distribution functions (CDF) $\mathbf{F}_s$ and $\mathbf{F}_o$ of the histograms $\mathbf{H}_s$ and $\mathbf{H}_o$ for the synthesized and the best-matched reference blocks respectively, the block distortion $D_b$ is defined as the \textit{Kolmogorov-Smirnov} (KS)\cite{lilliefors1967kolmogorov} distance between the two CDFs, \textit{i.e.} 
\begin{eqnarray}
\label{eq:blockD}
D_b = \max \limits_{j \in [1,L]} \left|\mathbf{F}_o(j)-\mathbf{F}_s(j)\right|
\end{eqnarray}
The overall normalized image distortion $d$ and the final image quality score $S$ are then computed as:
\begin{eqnarray}
d = \frac{1}{D_0}\sum_{b = 1}^{B} D_b, \; \text{and }
S = \frac{1}{1+d} 
\end{eqnarray}
where $D_0$ is a normalization constant. The score $S$ ranges in the interval $[0,1]$, where a lower distortion corresponding to a higher score and vice versa.

\subsection{A Proxy of 3DSwIM}

When approximating a contour in a pixel block, given a window of $K$ previous edges $\{e_{t}, \ldots, e_{t-K+1}\}$, it is desirable to evaluate the effects on aggregate rate and distortion of choosing an edge $e_{t+1}$ \textit{locally} without considering edges \textit{globally} outside the window. 
The complex definition of 3DSwIM, however, makes this difficult. 
When computing distortion between a target and a reference block, 3DSwIM identifies the maximum difference between the CDFs of the two corresponding histograms of wavelet coefficients.
Hence the effects on distortion due to changes in one edge $e_{t+1}$ are not known until all other edges are decided, which together determine the bin populations of wavelet coefficients in the histogram.
This inter-dependency among edges in a block makes it difficult to minimize distortion systematically without exhaustively searching through all possible edge sequences $\mathcal{E}$. 

In response, we propose a simple proxy to mimic 3DSwIM during contour approximation. 
Specifically, we first prove that the sum of local distortions (for individual pixel rows) is an upper bound of global 3DSwIM (for the entire pixel block). 
Thus, we can minimize the sum of local distortions to minimize the upper bound of the global distortion. 
Second, by assuming that the CDF of wavelet coefficients for a pixel row follows a certain model, we compute the local distortion as the maximum difference between two CDFs, which reduces to a simple function of respective model parameters. 
The sum of local distortions in a block is then used as a distortion proxy for contour approximation. 

\subsubsection{Local Distortion Upper Bound}
\label{sec:upper}

In 3DSwIM, the 1D wavelet detail coefficients for different pixel rows in an $N \times N$ block are collected and sorted into $L$ bins to construct a histogram.
$\mathbf{F}_s$ and $\mathbf{F}_o$ are the CDFs of the histograms for the synthesized and the best-matched reference blocks respectively. 
Suppose that instead we divide coefficients in each row into the same $L$ bins, and define CDFs for each row $i$ as $\mathbf{f}^i_s$ and $\mathbf{f}^i_o$, for the synthesized and best-matched blocks respectively. 
We see easily that $\mathbf{F}_o = \sum_{i=1}^N \mathbf{f}^i_o$ and $\mathbf{F}_s = \sum_{i=1}^N \mathbf{f}^i_s$. 
Thus the block distortion in (\ref{eq:blockD}) can be rewritten as:
\begin{eqnarray}
\label{eq:bound1}
D_b  =  \max \limits_{j \in [1,L]} \left|\sum_{i=1}^N \mathbf{f}^i_o (j) - \sum_{i=1}^N \mathbf{f}^i_s (j) \right|
\end{eqnarray}

To derive an upper bound for $D_b$ in (\ref{eq:bound1}), we see that
\begin{eqnarray}
\label{eq:bound2}
D_b & \leq &  \max \limits_{j \in [1,L]} \; \sum_{i=1}^N \big|\mathbf{f}^i_o (j) - \mathbf{f}^i_s (j)\big| \\
\label{eq:bound3}
& \leq & \sum_{i=1}^N  \; \max \limits_{j \in [1,L]} \big|\mathbf{f}^i_o (j) - \mathbf{f}^i_s (j)\big|  \\ \nonumber
& \triangleq & \sum_{i=1}^N D_b^i
\end{eqnarray}
where $D_b^i$ denotes the maximum difference between the CDFs of the coefficient histograms for pixel row $i$; we call this the \textit{row distortion}. 
We can now write our distortion proxy $\hat{D}_b$ that mimics 3DSwIM as the sum of individual row distortions, \textit{i.e.}
\begin{eqnarray}
\label{eq:proxyFinal}
\hat{D}_b = \sum_{i=1}^N D_b^i.
\end{eqnarray}

\subsubsection{A Model for Row Distortion}
\label{sec:MofD}

We now investigate how to compute row distortion $D^i_b$ efficiently. 
We first model the wavelet coefficients of a color pixel row using a \textit{Laplace distribution}. 
We then compute the row distortion as a function of the model parameters.

In \cite{van1999statistical}, Wouwer \textit{et al.} show that for a color image block with size $64 \times 64$, the histogram of the AC wavelet transform coefficients on each subband can be modelled by a generalized Gaussian density function:
\begin{eqnarray}
\label{eq:gaussianLaw}
f_{\sigma,\rho}(c) &=& \frac{\rho}{2\sigma\cdot \Gamma(\frac{1}{\rho})} \cdot e^{ -(\frac{|c|}{\sigma})^\rho }, \; c\in \mathbb{R}\\ \nonumber
\Gamma(x) &=& \int_{0}^{\infty} e^{-t}t^{x-1} dt,\; x > 0
\end{eqnarray}
where $\sigma$ is the variance, and $\rho$ is a shape parameter ($\rho = 2$ or $1$ for a Gaussian or Laplace distribution). 
Histogram $f_{\sigma,\rho}(c)$ can also be interpreted as the probability of a coefficient taking on value $c$. 

Inspired by \cite{van1999statistical}, in our work we assume also that the AC wavelet coefficients on a color pixel row follow the same distribution (\ref{eq:gaussianLaw}). 
For simplicity, we assume further that $\rho = 1$, \textit{i.e.}, the AC coefficients follow a Laplace distribution with parameter $\sigma$, where
\begin{eqnarray}
\label{eq:laplace}
f_\sigma(c) = \frac{1}{2\sigma} \cdot e^{-\frac{|c|}{\sigma}},\; c\in \mathbb{R}
\label{eq:pdfLaplace}
\end{eqnarray}
since $\Gamma(1) = 1$. 
In practice, given $M$ observed AC coefficients $\{c_1,\cdots,c_M\}$, the best estimate of model parameter $\sigma$ using the \textit{maximum likelihood estimation} (MLE)~\cite{johansen1990maximum} criteria is 
\begin{eqnarray}
\label{eq:sigma}
\sigma = \frac{1}{M} \sum_{i = 1} ^ M |c_i|
\end{eqnarray}
The derivation for (\ref{eq:sigma}) is in Appendix A. 

Assuming that the AC coefficients of a pixel row on the synthesized and its best-matched reference blocks are both Laplace distributions with respective parameters $\sigma_s$ and $\sigma_o$, we now compute the row distortion $D_b^i$ as a function of $\sigma_s$ and $\sigma_o$.
Since $D_b^i$ computes the maximum difference between the CDFs of two histograms, we first write the CDF of $f_\sigma(c)$ as: 
\begin{eqnarray}
F_\sigma(c) &=& \int_{-\infty}^c f_\sigma(x) \mathrm{d}x \\ \nonumber
&=& \left\{ \begin{array}{ll}
\frac{1}{2} \exp\left \{\frac{c}{\sigma} \right \}, & \; \text{if } c<0\\
1 - \frac{1}{2} \exp\left\{-\frac{c}{\sigma}\right\}, & \; \text{if } c \geq 0 \; .
\end{array} \right.
\end{eqnarray}
Then we define $g(c) = \left|F_{\sigma_o}(c) - F_{\sigma_s}(c)\right|$, and $D_b^i$ is equivalent to finding the maximum value of $g(c)$, $ c\in \mathbb{R}$. 

For simplicity, we define $\sigma_\mathtt{max} = \max \{\sigma_o,\sigma_s\} $ and $\sigma_\mathtt{min} = \min \{\sigma_o,\sigma_s\} $. 
When $\sigma_\mathtt{max} = \sigma_\mathtt{min}$, $g(c) = 0$. 
When $\sigma_\mathtt{max} > \sigma_\mathtt{min}$, ignoring the constant weight $1/2$, we get
\begin{eqnarray}
g(c) = \left \{ \begin{array}{ll}
 \exp \left \{\frac{c}{\sigma_\mathtt{max}}\right \} - \exp \left \{ \frac{c}{\sigma_\mathtt{min}} \right \} , & \; \text{if } c < 0 \\
\exp \left \{-\frac{c}{\sigma_\mathtt{max}}\right \} - \exp \left \{-\frac{c}{\sigma_\mathtt{min}}\right \}, & \; \text{if } c \geq 0  \; 
\end{array} \right.
\end{eqnarray}
The first derivative function of $g(c)$ becomes
\begin{eqnarray}
g'(c) = \left \{ \begin{array}{ll} \!\!\!
\frac{1}{\sigma_\mathtt{max}} \exp \left \{\frac{c}{\sigma_\mathtt{max}} \right \} - \frac{1}{\sigma_\mathtt{min}}  \exp \left \{\frac{c}{\sigma_\mathtt{min}} \right \},\!\!\! & \text{if } c < 0 \\
\!\!\! \frac{1}{\sigma_\mathtt{min}} \exp \left \{ \frac{-c}{\sigma_\mathtt{min}} \right \} -\frac{1}{\sigma_\mathtt{max}}\exp \left \{ \frac{-c}{\sigma_\mathtt{max}} \right \}, \!\!\! &\text{if } c \geq 0  
\end{array} \right.
\end{eqnarray}
When we set $g'(c) = 0$, we get the optimal $c$ as:
\begin{eqnarray}
c = \left \{
\begin{array}{ll}
\frac{\sigma_\mathtt{max} \sigma_\mathtt{min}}{\sigma_\mathtt{max}-\sigma_\mathtt{min}} \ln \frac{\sigma_\mathtt{min}}{\sigma_\mathtt{max}} \; \triangleq c^*, & \text{if } c < 0 \\
- c^*, & \text{if } c \geq 0  \; 
\end{array} \right.
\end{eqnarray}
We thus conclude that $g(c^*)$ is a maximum for $c < 0$ and $g(-c^*)$ is a maximum for $c \geq 0$. Since $g(c^*) = g(-c^*)$, the maximum value of $g(c)$ becomes
\begin{eqnarray}
g_\mathtt{max} = g(c^*) =  \left(\frac{\sigma_\mathtt{min}}{\sigma_\mathtt{max}}\right)^{\frac{\sigma_\mathtt{min}}{\sigma_\mathtt{max}-\sigma_\mathtt{min}}} - \left(\frac{\sigma_\mathtt{min}}{\sigma_\mathtt{max}}\right)^{\frac{\sigma_\mathtt{max}}{\sigma_\mathtt{max}-\sigma_\mathtt{min}}}. 
\end{eqnarray}


In summary, for a pixel row in the synthesized and best-matched blocks with respective histograms $f_{\sigma_s}$ and $f_{\sigma_o}$, the row distortion $D_b^i$ 
can be computed as:
\begin{eqnarray}
\label{eq:reformulate}
D_b^i & = & \langle f_{\sigma_o}, f_{\sigma_s} \rangle \\ \nonumber
& = & \left\{ \begin{array}{ll} \!\!\!
(\frac{\sigma_\mathtt{min}}{\sigma_\mathtt{max}})^{\frac{\sigma_\mathtt{min}}{\sigma_\mathtt{max}-\sigma_\mathtt{min}}} - (\frac{\sigma_\mathtt{min}}{\sigma_\mathtt{max}})^{\frac{\sigma_\mathtt{max}}{\sigma_\mathtt{max}-\sigma_\mathtt{min}}},  & \text{if } \sigma_\mathtt{max} > \sigma_\mathtt{min}  \\ \!\!\!
0,  & \text{if } \sigma_\mathtt{max} = \sigma_\mathtt{min}
\end{array}  \right.
\end{eqnarray}
where $\sigma_\mathtt{max} = \max \{\sigma_o,\sigma_s\} $ and $\sigma_\mathtt{min} = \min \{\sigma_o,\sigma_s\} $.

Together with (\ref{eq:proxyFinal}), our proposed distortion proxy that mimics 3DSwIM becomes
\begin{eqnarray}
\label{eq:proxy}
\hat{D}_b = \sum_{i=1}^N  \langle f_{\sigma_o^i}, f_{\sigma_s^i} \rangle.
\end{eqnarray}
where $\sigma_s^i$ and $\sigma_o^i$ are the respective Laplace distribution parameters for coefficients on the pixel row $i$ for the synthesized and best-matched blocks. 

\begin{figure}[htb]
\centering
\centerline{\includegraphics[width=5.8cm]{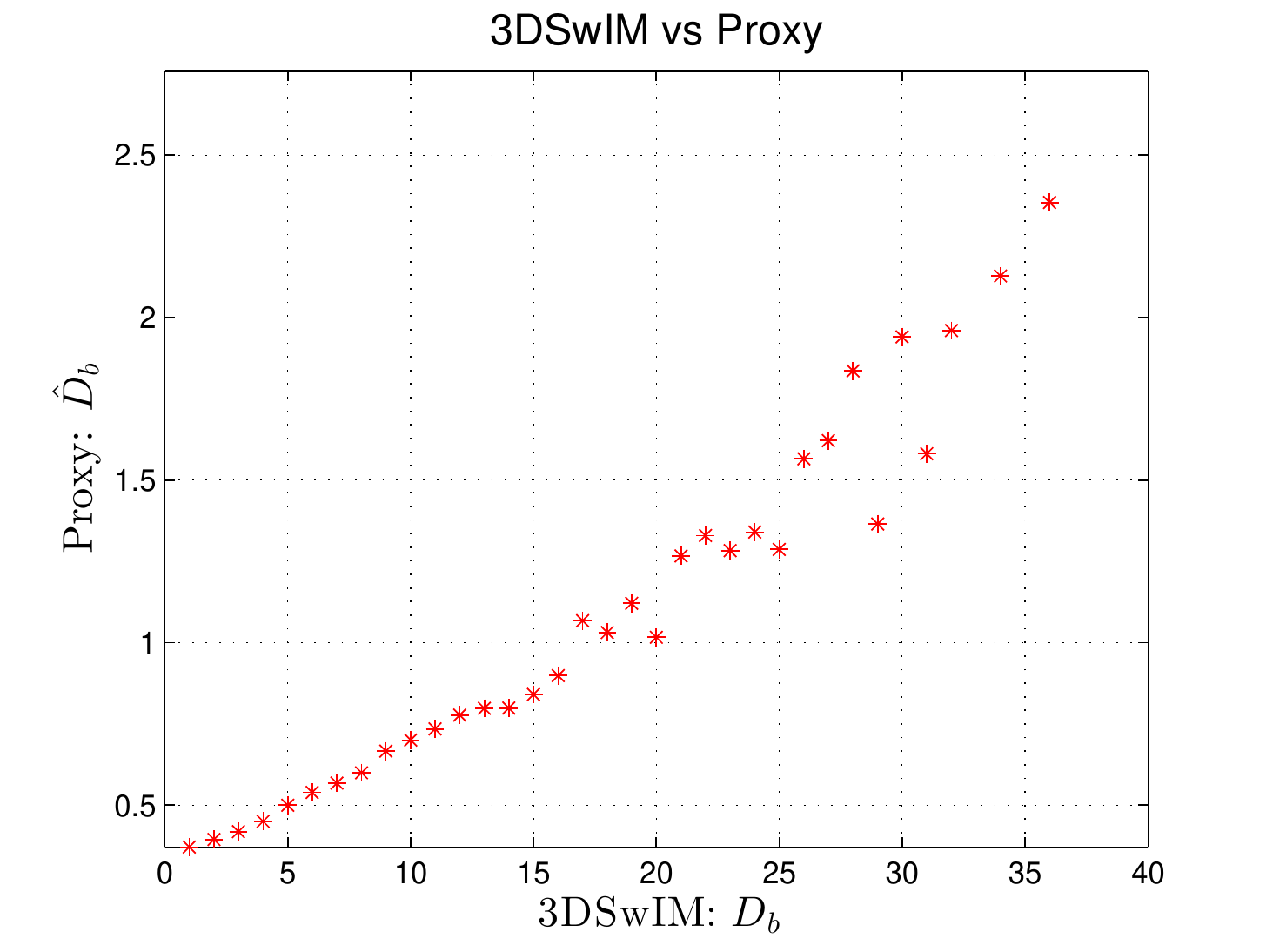}}
\caption{$x$-axis: block distortion calculated by 3DSwIM; $y$-axis: mean value of distortion calculated by our proposed proxy}
\label{fig:3DVSproxy}
\end{figure}

To verify the accuracy of our distortion proxy, we tested 17 image sequences from the Middleburry dataset\footnote{http://vision.middlebury.edu/stereo/data/}. Virtual views are synthesized by a simple implementation of 3D warping~\cite{tian09} and each image is divided into around 600 blocks with size $16 \times 16$ for distortion computation. The result is shown in Fig.\;\ref{fig:3DVSproxy}, in which the $x$-axis is $D_b$ by 3DSwIM and the $y$-axis is the mean value of our proxy $\hat{D}_b$ in the corresponding blocks. 
This result clearly shows a positive linear trend between $D_b$ and $\hat{D}_b$, demonstrating the effective approximation of our proposed distortion proxy.

\section{Object Contour Approximation}
\label{sec:app}
Given the AEC-based contour coding and our proposed distortion proxy discussed in Sections~\ref{sec:rate} and~\ref{sec:dist} respectively, we now discuss how to approximate a detected contour.

\subsection{Dividing contour into segments}

When approximating a contour detected in a depth image,
we first divide the contour into \textit{segments}, where each segment is composed of edges with only two \textit{non-opposite} directions:  $\{ \downarrow,\rightarrow \}$, $\{ \downarrow,\leftarrow \}$, $\{ \uparrow,\leftarrow \}$ or $\{\uparrow,\rightarrow \}$. We then approximate each segment separately. 
We require each approximated segment to start and end at the same locations as the original segment. 
Given that edges in a segment can only take non-opposite directions, it implies that the length $T$ of the approximated segment must be the same as the original. 
Thus the search space of candidates is only $T \choose V$, where $V$, $V \leq T$, is the number of vertical edges ($\downarrow$ or $\uparrow$) in the original segment. 
To efficiently search for an RD-optimal approximated segment within this space, we propose to use a DP algorithm. 
Finally, the approximated segments are combined to form an approximated contour.

\subsection{DP algorithm to approximate one segment}

\subsubsection{Efficient computation of the distortion proxy}

We first discuss how to efficiently compute our distortion proxy during segment approximation. 
Our distortion proxy Eq. (\ref{eq:proxy}) is computed by comparing a pixel block in a synthesized view image with a best-matched block in the reference image.
To avoid laboriously augmenting the input color / depth image pair and synthesizing a virtual view image to compute distortion for each segment approximation, we propose to use \textit{only the input color image} and a simple \textit{shifting window} procedure to compute our distortion proxy. 
We describe this next.

\begin{figure}[htb]
\centering
\centerline{\includegraphics[bb=0 0 630 270, width=8.5cm]{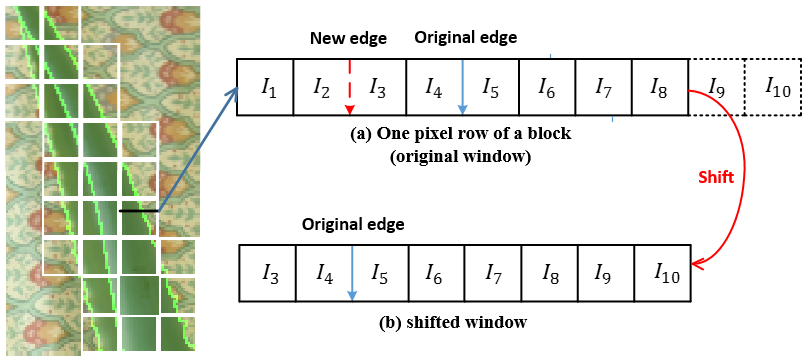}}
\caption{ An example of shifting window. The detected contours are shown in green in the color image. The white blocks contain part of contour segments. The original edge crossing one pixel row (black) in a block is shifted \textit{left} by 2 pixels to a new edge. We shift the original window \textit{right} by 2 pixels and identify the shifted window.}
\label{fig:colorShift}
\end{figure}

We first divide the input color image into non-overlapping  $N \times N$ pixel blocks.
The block distortion---the sum of row distortions according to our proxy of Eq. (\ref{eq:proxy})---is computed for blocks that contain parts of the segment. 
See Fig. \ref{fig:colorShift} for an illustration of blocks of a color image containing contour segments (in green). 
When a segment is approximated and altered, \textit{vertical edges} in the original segment are horizontally shifted.  
As an example, in Fig.\;\ref{fig:2DEdges}(a), the vertical edge $e^o_3$ in the original blue segment is shifted left by one pixel to $e_4$ in the approximated red segment. 

At a given pixel row, assuming that the contained vertical edge has horizontally shifted, we compute the resulting row distortion using a shifting window procedure as follows. 
Suppose that an original vertical edge starting from 2D coordinate $(p^o,q^o)$ (specifying respective row and column indices) is horizontally shifted by $k$ pixels (positive / negative $k$ means shifting to the right / left) to a new vertical edge with starting point $(p^o,q)$, where $k = q - q^o$. 
Delimiting the pixel row in the original block with a $N$-pixel window, we shift the window 
by $-k$ pixels to identify a new set of $N$ pixels that represent the pixels in the corresponding block of the synthesized view after segment modification.

An example of shifting window is illustrated in Fig.\;\ref{fig:colorShift}. 
One pixel row (black) in an original $8 \times 8$ block contains the pixel set $\mathbf{u} = \{I_1,\cdots,I_8\}$. 
During segment approximation, the original vertical edge between pixels $I_4$ and $I_5$ shifts \textit{left} by 2 pixels $(k = -2)$ to a new edge. 
As a result, we shift the delimiting window \textit{right} by 2 pixels, and identify a different set of eight pixels, $\mathbf{v} = \{I_3,\cdots,I_{10}\}$. 
We see that the original edge in the shifted window is located two pixels from the left, and the new edge in the original window is also located two pixels from the left.
Because of this alignment, we can simply use window $\mathbf{v}$ as a representation of the pixel row after segment approximation, instead of the original pixels $\mathbf{u}$ plus image augmentation due to the vertical edge shift.

We can now compute the row distortion using (\ref{eq:reformulate}) using two sets of $N$ pixels delimited by the window before and after the shift operation. 
Because 3DSwIM only searches for a best-matched reference block within a window of size $2W$ pixels, when $|k| > W$, pixels in the shifted window no longer well represents the best-matched pixel row. 
We thus set the distortion to infinity in this case to signal a violation. 
We can now write the row distortion $d_{p^o}(q^o,q)$ induced by a horizontally shifted vertical edge from start point $(p^o,q^o)$ to $(p^o,q)$ for the $p^o$-th pixel row as
\begin{eqnarray}
d_{p^o}(q^o,q) = \left\{ \begin{array}{ll}
\langle f_{\sigma_{\textbf{u}}}, f_{\sigma_{\textbf{v}}} \rangle, & \text{if } \; |q-q^o| \leq W \\
\infty, & \text{O.W.}
\end{array} \right.
\label{eq:effRowDist}
\end{eqnarray}
where $\sigma_{\textbf{u}}$ and $\sigma_{\textbf{v}}$ are the respective Laplace distribution parameters computed using pixels in the original and the shifted windows $\mathbf{u}$ and $\mathbf{v}$.
Continuing with the example in Fig.\;\ref{fig:2DEdges}(a), $d_2(3,2)$ is the distortion of edge $e_3^o$ (starting from $(2,3)$) being shifted to edge $e_4$ (starting from $(2,2)$). 

\subsubsection{DP algorithm}

We next design a DP algorithm to approximate a contour segment composed of edges with directions $\{ \downarrow, \leftarrow \}$. 
The algorithms for segments composed of edges with two different non-opposite direction pairs are similar, and thus are left out for brevity. 
We first define distortion and rate terms. 
Then we define a DP recursive equation based on Lagrangian relaxation. 

\paragraph{Problem formulation}

\begin{figure}[htb]
\hspace{0.1in}
\begin{minipage}[b]{.23\textwidth}
  \centering
  \centerline{\includegraphics[width=4.3cm]{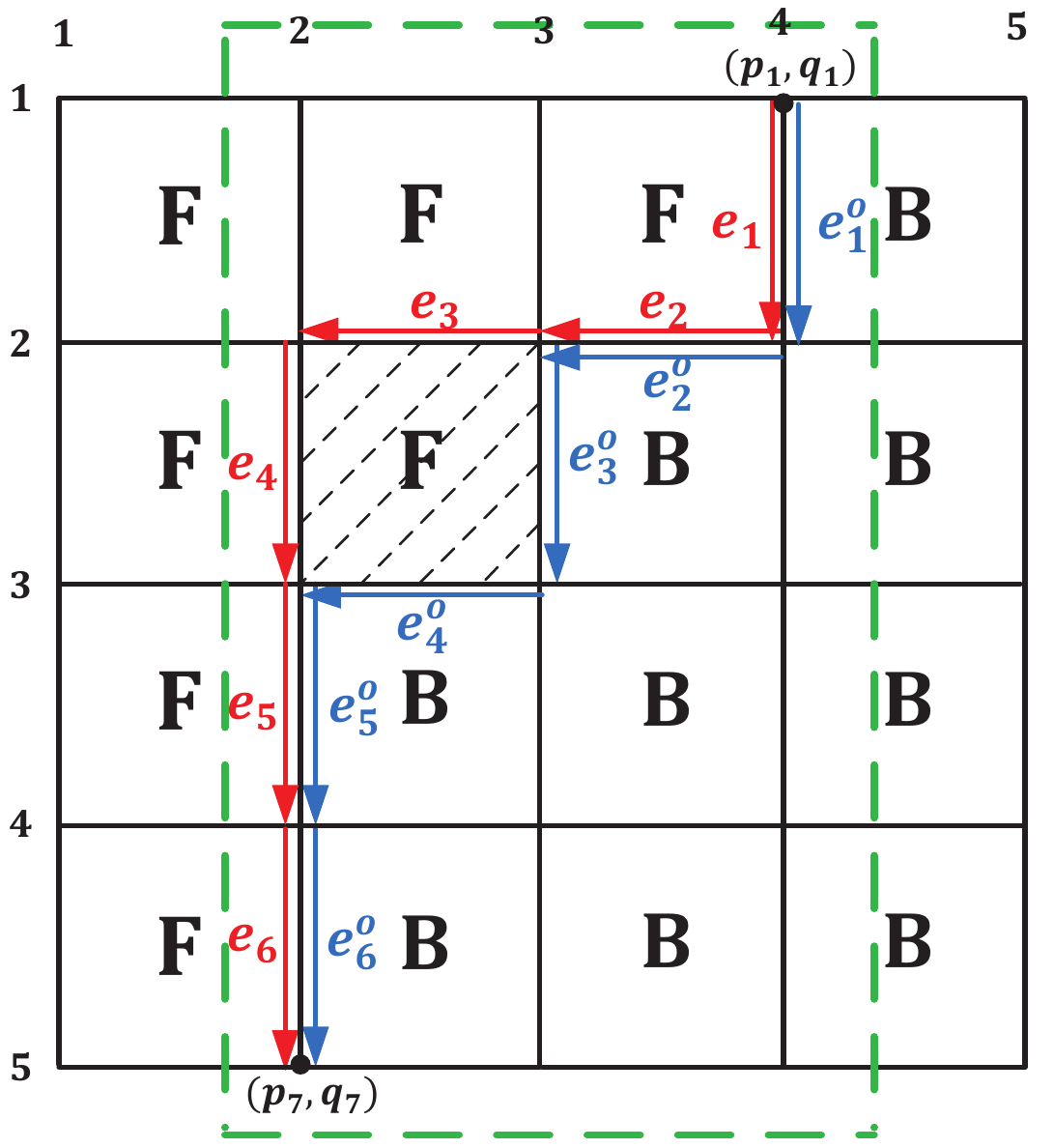}}
  \centerline{(a)}\medskip
\end{minipage}
\begin{minipage}[b]{.23\textwidth}
  \centering
  \centerline{\includegraphics[width=3.7cm]{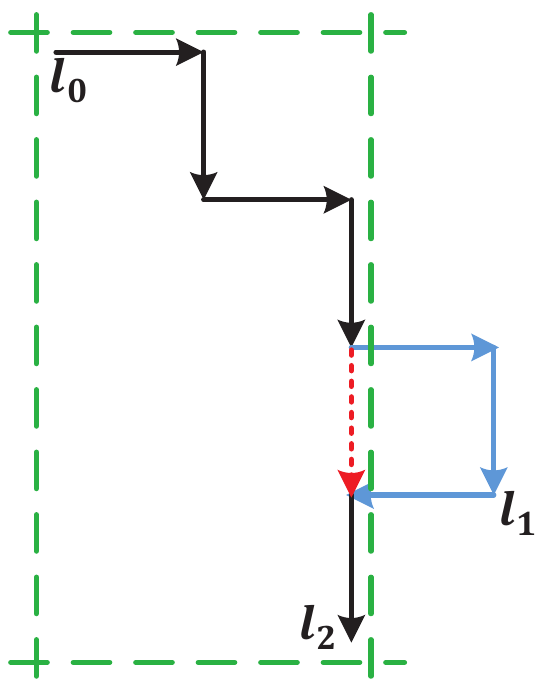}}
  \centerline{(b)}\medskip
\end{minipage}
\caption{(a) An example of segment approximating. The original segment $\mathcal{E}^o = \{ e_1^o, \cdots, e^o_6 \}$ separating foreground (F) and background (B), where the approximated segment $\mathcal{E} =\{e_1, \cdots, e_6 \}$.
The vertical edge $e_4$ is shifted by edge $e_3^o$. The pixel with the top left corner $(2,2)$ need to be altered after approximation. (b) An example of segment merging.}
\label{fig:2DEdges}
\end{figure}

Denote by $\mathcal{E}^o$ and $\mathcal{E}$ the original and approximated segments respectively, and by $T$ the length of $\mathcal{E}^o$. 
Further, we define $(p_1,q_1)$ as the 2D coordinate of the first edge's start point. Given $V$ vertical edges with direction $\downarrow$ in $\mathcal{E}^o$, the coordinate of the last edge's end point can be computed as $(p_{T+1},q_{T+1}) = (p_1 + V, q_1 - (T-V))$.  
Given our constraint that $\mathcal{E}$ must start and end at the same locations, the search space $\mathcal{S}$ for $\mathcal{E}$ is restricted to the rectangle region with opposite corners at $(p_1,q_1)$ and $(p_{T+1},q_{T+1})$. 
As an example, the original blue segment in Fig.\;\ref{fig:2DEdges}(a) has length $T=6$ and $V = 4$, with start point $(p_1,q_1) = (1,4)$ and end point $(p_7,q_7) = (5,2)$. 
The search space is the green rectangle region.

Given that the row distortion is induced by a horizontally shifted vertical edge,
the total distortion induced by an approximated segment is the sum of row distortions induced by all shifted vertical edges. 
Using the defined row distortion $d_{p^o}(q^o,q)$ for the $p^o$-th row, given an approximated segment $\mathcal{E} = \{e_T,\cdots,e_1\}$ with two non-opposite directions $\{\downarrow,\leftarrow \}$,  
the total distortion for approximating $\mathcal{E}^o$ with $\mathcal{E}$ is:
\begin{equation}
D(\mathcal{E}, \mathcal{E}^o) = \sum \limits_{ e_i \in \mathcal{E} | e_i = \downarrow } d_{p^o_i}(q^o_i,q_i)
\end{equation}
where $(p^o_i,q_i)$ is the 2D coordinate of edge $e_i$'s starting point in $\mathcal{E}$, and $q^o_i$ is the column index of a vertical edge on $\mathcal{E}^o$ crossing the $p^o_i$-th pixel row. 

To approximate a segment $\mathcal{E}^o$, we solve the following RD cost function:
\begin{eqnarray}
\label{eq:object}
\min \limits_{\mathcal{E}} \; D(\mathcal{E}, \mathcal{E}^o) + \lambda R(\mathcal{E})
\end{eqnarray}
where $R(\mathcal{E})$ is the coding rate of segment $\mathcal{E}$ using AEC edge coding (Section~\ref{sec:rate}). 
Given a window of $K$ previous edges $\mathbf{s}_t=\{e_{t-1},\cdots,e_{t-K}\}$ and the computed probability $P(e_{t} | \mathbf{s}_t)$ based on AEC, we use the entropy to estimate the coding bits of edge $e_{t}$, \textit{i.e.}, $r(e_{t} | \mathbf{s}_t) = - \log_2 P(e_{t} | \mathbf{s}_t) $ 
\footnote{Given a segment, when $t \leq K$, we define $\mathbf{s}_t$ as the combination of the first $t-1$ edges in the current segment and the last $K-t+1$ edges in the previous segment. As to the the first segment in a contour, when $t \leq K$, we define $\mathbf{s}_t = e_{t-1} \cup \mathbf{s}_{t-1}$ with $\mathbf{s}_1 = \emptyset$, and $P(e_{1} | \mathbf{s}_1) = 1/4$ since $e_1 \in \mathcal{A}^o$ with $|\mathcal{A}^o| = 4$, and $P(e_{t} | \mathbf{s}_t) = 1/3$ for $t>1$, since $e_t \in \mathcal{A}$ with $\mathcal{|A|} = 3$.}.
%
Thus the total segment coding rate becomes
\begin{equation}
\label{eq:rate}
R(\mathcal{E}) =  \sum \limits_{t=1}^{T} r(e_{t} | \mathbf{s}_{t})
\end{equation}

\paragraph{DP algorithm development}

We propose a DP algorithm to solve (\ref{eq:object}). 
First, we define $J\left(\mathbf{s}_t,p_{t},q_{t}\right)$ as the recursive RD cost of a partial segment from the $t$-th edge to the last edge, given that the $K$ previous edges are $\mathbf{s}_t \in \{\downarrow,\leftarrow \}^K$, and $(p_{t},q_{t})$ is the coordinate of the $t$-th edge $e_t$'s starting point.  
$J\left(\mathbf{s}_t,p_{t},q_{t}\right)$ can be recursively solved by considering each candidate $e_t$ inside the search space $\mathcal{S}$ (the rectangle box):
\begin{align}
\label{eq:DPalg}
 & J\left(\mathbf{s}_t, p_{t},q_{t}\right) =  \min \limits_{e_{t} \in \big\{ \{\downarrow,\leftarrow\} | (p_{t+1},q_{t+1}) \in \mathcal{S} \big \} } \Big\{ 
 \lambda \cdot r(e_{t} | \mathbf{s}_t)   \\ \nonumber
& + d_{p_t^o}(q^o_t,q_t ) \cdot \delta_\downarrow + J\left( \mathbf{s}_{t+1}, p_{t+1},q_{t+1} \right) \cdot \delta(p_{t+1},q_{t+1}) \Big\}
\end{align}
where $p_t^o = p_t$ and $\delta_\downarrow $ is a binary indicator that equals to 1 if $e_{t} = \downarrow$ and 0 otherwise. The value $\mathbf{s}_{t+1}$ is updated by combining $e_{t}$ with the first $K-1$ elements in $\mathbf{s}_t$. 
The pixel position $(p_{t+1},q_{t+1})$ equals to $(p_t+1,q_{t})$ or $(p_t,q_t-1)$ if $e_{t} = \downarrow$ or $e_{t} = \leftarrow $, respectively.  Finally $\delta(p,q)$ indicates the termination condition, which is
\begin{eqnarray}
\delta(p,q) = \left\{ \begin{array}{ll}
0, & \text{if } \; (p,q) = (p_{T+1},q_{T+1}) \\
1, & \text{o.w.}
\end{array} \right.
\end{eqnarray}

Eq. (\ref{eq:DPalg}) is considered a DP algorithm because there are overlapping sub-problems during recursion. 
Each time a sub-problem with the same argument is solved using (\ref{eq:DPalg}), the solution is stored as an entry into a DP table. 
Next time the same sub-problem is called, the previously computed entry is simply retrieved from the DP table and returned. 

Starting from the first edge, the recursive call 
results in the minimum cost of the segment approximation. The corresponding segment $\mathcal{E}^* = \text{arg} \min J\big(\mathbf{s}_{1},p_{1},q_{1}\big)$ is the new approximated segment.

\subsubsection{Complexity of DP algorithm}


The complexity of a DP algorithm is upper-bounded by the size of DP table times the complexity to compute each table entry. 
The size of the DP table can be bounded as follows. 
The argument $\mathbf{s}_t$ in $J(\mathbf{s}_t, p_t, q_t)$ can take on $O(2^K)$ values.
On the other hand, the argument $(p_t,q_t)$---the starting point of $e_t$---can take on $O(V \times (T-V))$ locations in search space $\mathcal{S}$. 
Thus the DP table size is $O(2^K V(T-V))$.  
To compute each DP table entry using Eq. (\ref{eq:DPalg}), a maximum of two choices for $e_t$ need to be tried, and each choice requires a sum of three terms. 
The second term involves row distortion $d_{p_t^o}(q^o_t,q_t )$, which itself can be computed and stored into a DP table of size $O(V W)$, each entry computed in $O(N^2)$.
Thus the complexity of computing $d_{p_t^o}(q^o_t,q_t )$ is $O(V W N^2)$.
Thus the overall complexity of (\ref{eq:DPalg}) is $O(2^K V(T-V) + V W N^2)$.  

In our experiments, we set $K = 3$ as done in \cite{daribo2012icip}, $W = 10$ and $N = 16$. The number $T-V$ of horizontal edges in a segment is usually smaller than $N^2$, thus the overall complexity of the algorithm can be simplified to $O(V W N^2)$.

%

\subsection{Greedy Segment Merging}

We next discuss how two consecutive segments can be merged into one to further reduce edge coding cost. Given the start point $l_0$ of the first segment and the end point $l_2$ of the second, 
we first delimit the feasible space of edges by the rectangle with opposing corners at $l_0$ and $l_2$. 
The original edges which are outside this feasible space are first projected to the closest side of the rectangle, so that edges in the feasible space become a new segment. This edge projection leads to edge shifting and thus induce distortion; we call this the \textit{merge distortion}. 
We then execute the aforementioned DP algorithm on this new simplified segment to get a new minimum cost. If this cost plus the merge distortion is smaller than the cost sum of the original two segments, we merge the two original segments to this new simplified segment.  

As illustrated in Fig.\;\ref{fig:2DEdges}(b), a segment with directions $\{\rightarrow,\downarrow\}$ starts from point $l_0$ and ends at $l_1$. Another segment with directions $\{\leftarrow,\downarrow\}$ exists between $l_1$ to $l_2$. The feasible space is marked as the green rectangle region.
The original blue edges outside the feasible space are projected to the dash red edge, such that the new segment from $l_0$ to $l_2$ only contains two directions $\{\rightarrow,\downarrow\}$.

The procedure of approximating one contour using our proposed DP algorithm and greedy segment merging method is summarized in Algorithm \ref{alg:approximation}.


\begin{algorithm} 
\caption{Depth Contour Approximation}
\label{alg:approximation}
\begin{algorithmic}[1]
\STATE {Divide a contour into $M$ segments;}
\STATE {Given a $\lambda$, execute Eq. (\ref{eq:DPalg}) on each segment to get the RD cost;}
\STATE {Greedily merge two neighbouring segments until there is no cost decrease;}
\STATE {Finally get $M'$ segments, where $M' \leq M$.}
\end{algorithmic}
\end{algorithm}


\section{Image Coding with Approximated Contours}
\label{sec:coding}
Given a set of approximated contours in an image that are coded losselessly using AEC \cite{daribo2012icip}, we now describe how to code depth and color images at the encoder. 
We also discuss inter-view consistency issues.

\subsection{Depth and Color Images Augmentation}

Approximating object contours means modifying the geometry of a 3D scene. 
Accordingly, both depth and color images from the same viewpoint need to be augmented to be consistent with the new geometry. 
Specifically, for the depth image, we exchange foreground and background depth pixel values across a modified edge, so that all foreground (background) depth values fall on the same side of an edge. In Fig.\;\ref{fig:2DEdges}(a), the pixel with the top left corner index $(2,2)$ need to be replaced by a background pixel.

For the color image, pixels corresponding to the altered pixels in the depth image are removed and labeled as holes. 
An image inpainting algorithm~\cite{criminisi04} is then used to inpaint the identified holes, with a constraint that only existing background (foreground) pixels can be used to complete holes in the background (foreground). 

\begin{figure}[htb]
\begin{minipage}[b]{.48\linewidth}
  \centering
  \centerline{\includegraphics[bb=0 0 164 60, width=4.1cm]{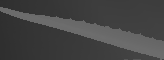}}
 \centerline{(a)}\medskip
\end{minipage}
\hfill
\begin{minipage}[b]{.48\linewidth}
  \centering
  \centerline{\includegraphics[bb=0 0 164 60, width=4.1cm]{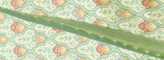}}
  \centerline{(e)}\medskip
\end{minipage}
\hfill
\begin{minipage}[b]{.48\linewidth}
  \centering
  \centerline{\includegraphics[bb=0 0 164 60, width=4.1cm]{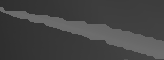}}
  \centerline{(b)}\medskip
\end{minipage}
\hfill
\begin{minipage}[b]{.48\linewidth}
  \centering
  \centerline{\includegraphics[bb=0 0 164 60, width=4.1cm]{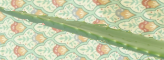}}
  \centerline{(f)}\medskip
\end{minipage}
\hfill
\begin{minipage}[b]{.48\linewidth}
  \centering
  \centerline{\includegraphics[bb=0 0 164 60, width=4.1cm]{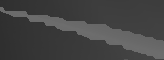}}
  \centerline{(c)}\medskip
\end{minipage}
\hfill
\begin{minipage}[b]{.48\linewidth}
  \centering
  \centerline{\includegraphics[bb=0 0 164 60, width=4.1cm]{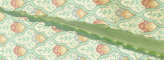}}
  \centerline{(g)}\medskip
\end{minipage}
\hfill
\begin{minipage}[b]{.48\linewidth}
  \centering
  \centerline{\includegraphics[bb=0 0 164 60, width=4.1cm]{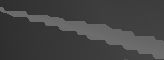}}
  \centerline{(d)}\medskip
\end{minipage}
\hfill
\begin{minipage}[b]{.48\linewidth}
  \centering
  \centerline{\includegraphics[bb=0 0 164 60, width=4.1cm]{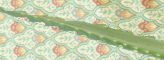}}
  \centerline{(h)}\medskip
\end{minipage}
\caption{Example for contour approximation and image value alteration.}
\label{fig:DepApp}
\end{figure}

Fig.\;\ref{fig:DepApp} shows an example of contour approximation and image value augmentation for one depth and color image pair. The two images (a) and (e) are part of the original depth and color image pair for \texttt{Aloe}. Depth contours are approximated by our proposed method with an increasing value of $\lambda$ for images from (b) to (d). We alter depth pixel values, and the PWS characteristic is still preserved in the approximated depth images. Images (f) to (h) are the corresponding inpainted color images, which looks natural along its neighboring foreground / background regions. The geometry structure in the altered depth and color image pairs are consistent with the approximated contours.

\subsection{Inter-view Consistency}

Given that the encoder needs to encode two color-plus-depth image pairs of the same 3D scene (e.g., left and right views) for synthesis of an intermediate virtual view at the decoder, we can perform contour approximation in an inter-view consistent manner as follows.


First, we approximate object contours of the left view and augment the corresponding depth and color images. 
We then project the approximated and modified left depth image to the right viewpoint via DIBR. 
With the projected depth image, we augment the original right depth and color image pair accordingly, such that the approximation on the left view is mapped to the right view. 
Finally, we approximate the augmented right depth and color images. To penalize the inter-view inconsistency when approximation happens on the right view, we add a penalty term to the row distortion (\ref{eq:effRowDist}), \textit{i.e.} we define the row distortion for right view approximation as
\begin{eqnarray}
d'_{p^o}(q^o,q) = d_{p^o}(q^o,q) + \rho |q - q^o|^2
\end{eqnarray}
where $|q - q^o|$ takes an additional meaning that the shifted edge on the augmented right view is now inconsistent with the approximated left view. The weight parameter $\rho$ is set very large to penalize the inter-view inconsistency. By doing so, the depth and color image pair on the right view is very likely consistent with that of the left view after contour approximation. 


\subsection{Edge-adaptive Image Coding}

The remaining task is to code the depth and color images, which have been modified accordingly to our contour approximation. 

We first describe how edge-adaptive GFT~\cite{hu2015tip} can be used to code the altered depth image.
For each pixel block in the altered depth image, a 4-connected graph is constructed, where each node corresponds to a pixel. The edge weight between two neighbouring pixels is defined based on their weak or strong correlations.
Given the constructed graph, the transform is the eigen-matrix of the graph Laplacian (see~\cite{hu2015tip} for details). 
The edge weight assignment (deducible from the encoded contours) means filtering across sharp boundaries are avoided, preventing blurring of edges.

Since our work is mainly about approximating object contours, here we focus on how contour approximation influences the depth coding rate. 
For color images, we code them with a state-of-the-art image coding method---HEVC HM 15.0~\cite{sullivan2012overview}.

\section{Experimentation}
\label{sec:results}
\subsection{Experimental Setup}

We perform extensive experiments to validate the effectiveness of our proposed contour approximation method using texture-plus-depth image sequences from the Middlebury dataset, \texttt{Undo Dancer} and \texttt{GFTly}~\cite{aflaki2011undo}. 
View synthesis for \texttt{Undo Dancer} and \texttt{GTFly} is performed using VSRS software\footnote{wg11.sc29.org/svn/repos/MPEG-
4/test/trunk/3D/view\_synthesis/VSRS}. 
Due to the lack of camera information, DIBR for the Middlebury sequences is performed using a simple implementation of 3D warping~\cite{tian09}.
 
For each image sequence, we first approximate the depth and color image pair with different values of $\lambda$ based on our proposed approximation method. 
We then deploy the GFT based edge-adaptive depth coding \cite{hu2015tip} to code the approximated and altered depth image using different quantization parameters (QP), where the approximated contours are losslessly coded as SI. 
The approximated and inpainted color image from the same viewpoint is compressed by HEVC intra HM 15.0~\cite{sullivan2012overview}.
Depth and color image pairs from different viewpoints but approximated by the same $\lambda$ are then transmitted to the decoder for virtual view synthesis.
The ground truth (reference image) is synthesized by the original depth and color images (no approximation and compression) for distortion computation. 
3DSwIM with block size $16 \times 16$ is used to evaluate the quality of the synthesized views, where a higher score means a better quality. 
As an additional metric, we also evaluate synthesized view quality with PSNR.
For the Middleburry sequences, we encode views 1 and view 5, and synthesize the intermediate virtual views $2$ to $4$. For \texttt{Undo Dancer} and \texttt{GFTly} sequences, views 1 and view 9 are encoded and the intermediate views $2$ to $8$ are synthesized.  
The average distortion of all the intermediate views is taken to evaluate the contour approximation performance.
For each image sequence, the convex hull of all operational points represents the rate-score (RS) or RD performance of our proposed method.

\begin{figure*}[t]
\begin{minipage}[b]{.23\textwidth}
  \centering
  \centerline{\includegraphics[width=4.6cm]{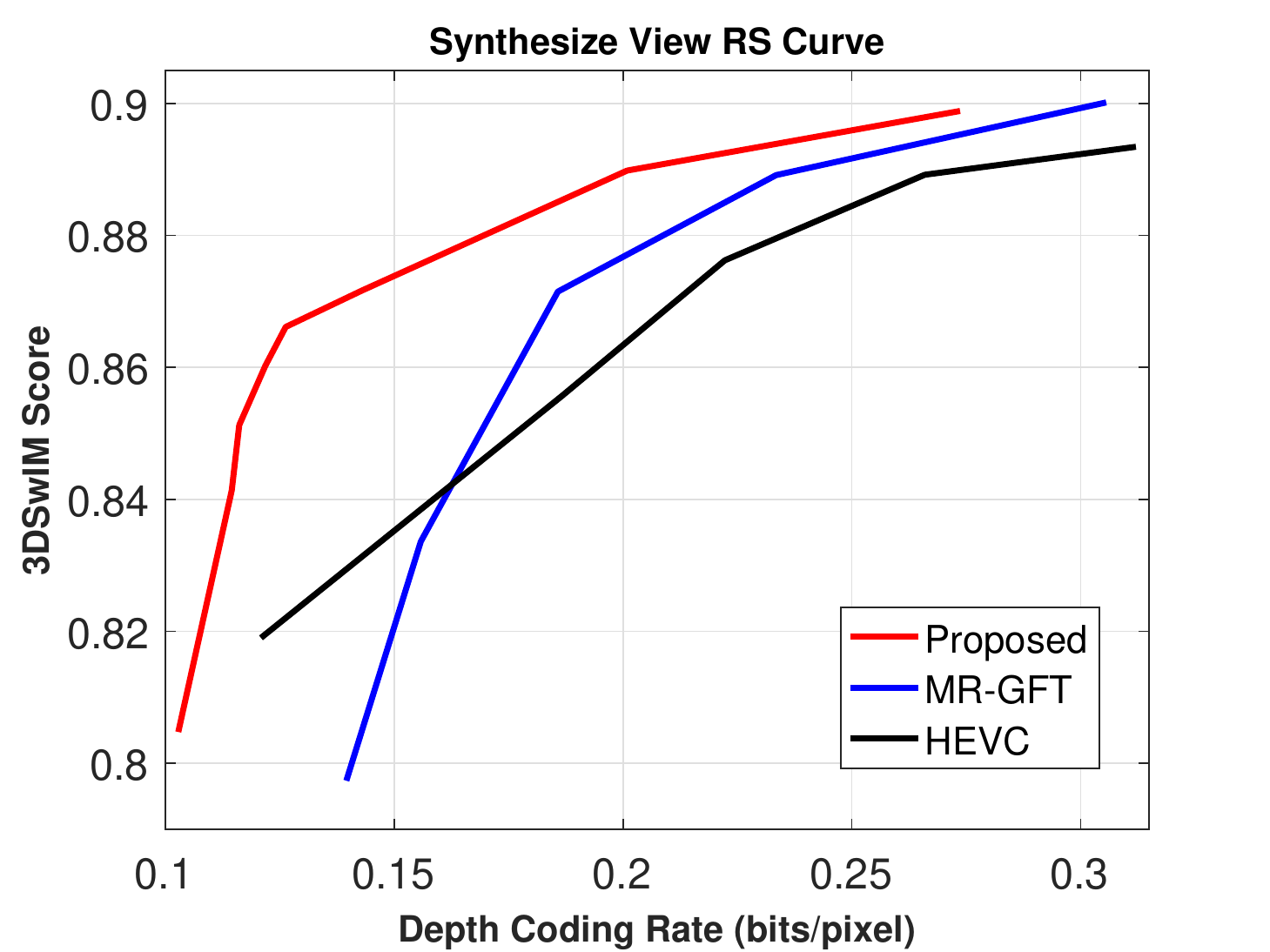}}
  \centerline{(a) \textrm{cones}}\medskip
\end{minipage}
\hfill
\begin{minipage}[b]{.23\textwidth}
  \centering
  \centerline{\includegraphics[width=4.6cm]{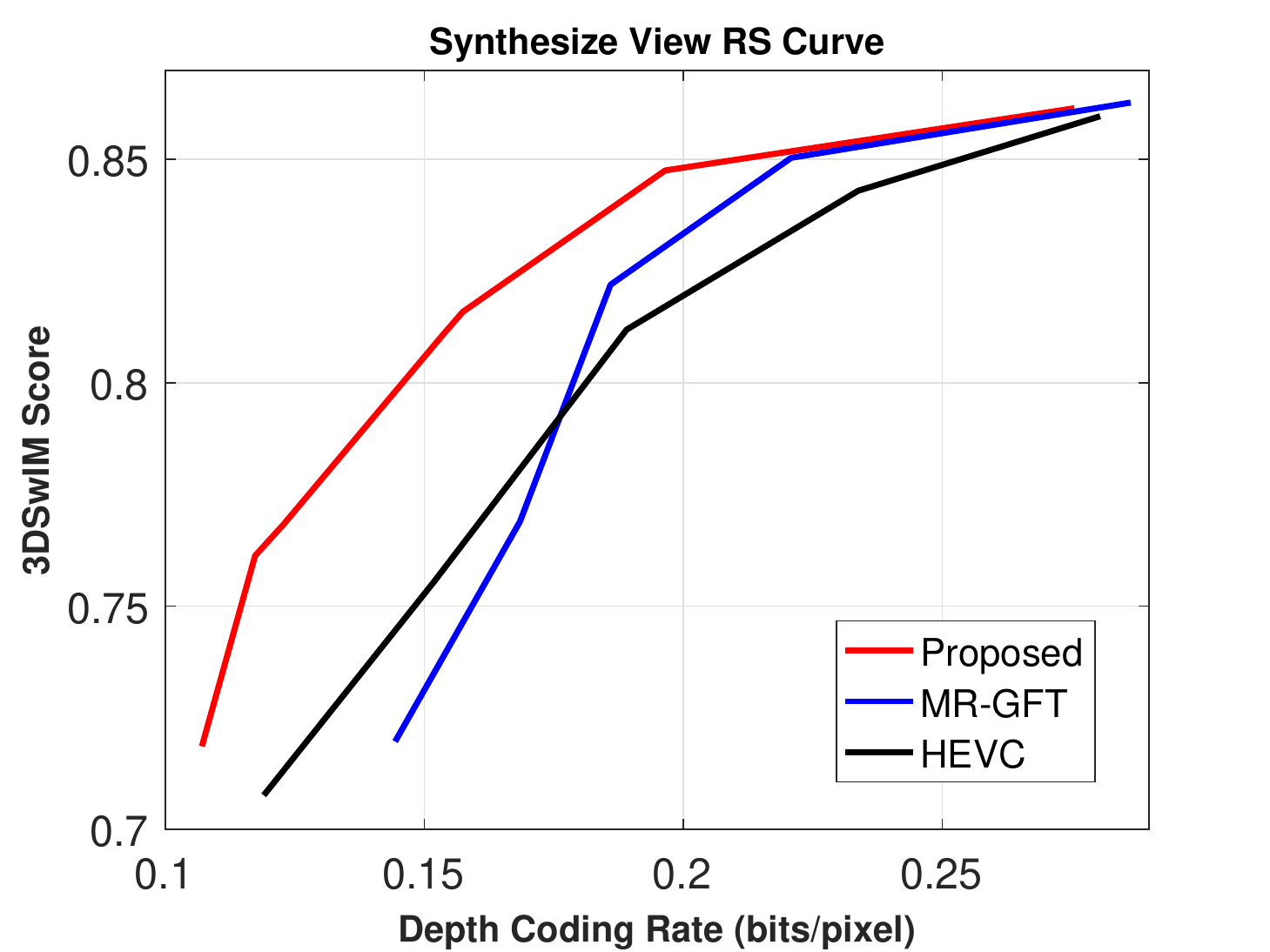}}
  \centerline{(b) \textrm{Moebius}}\medskip
\end{minipage}
\hfill
\begin{minipage}[b]{.23\textwidth}
  \centering
  \centerline{\includegraphics[width=4.6cm]{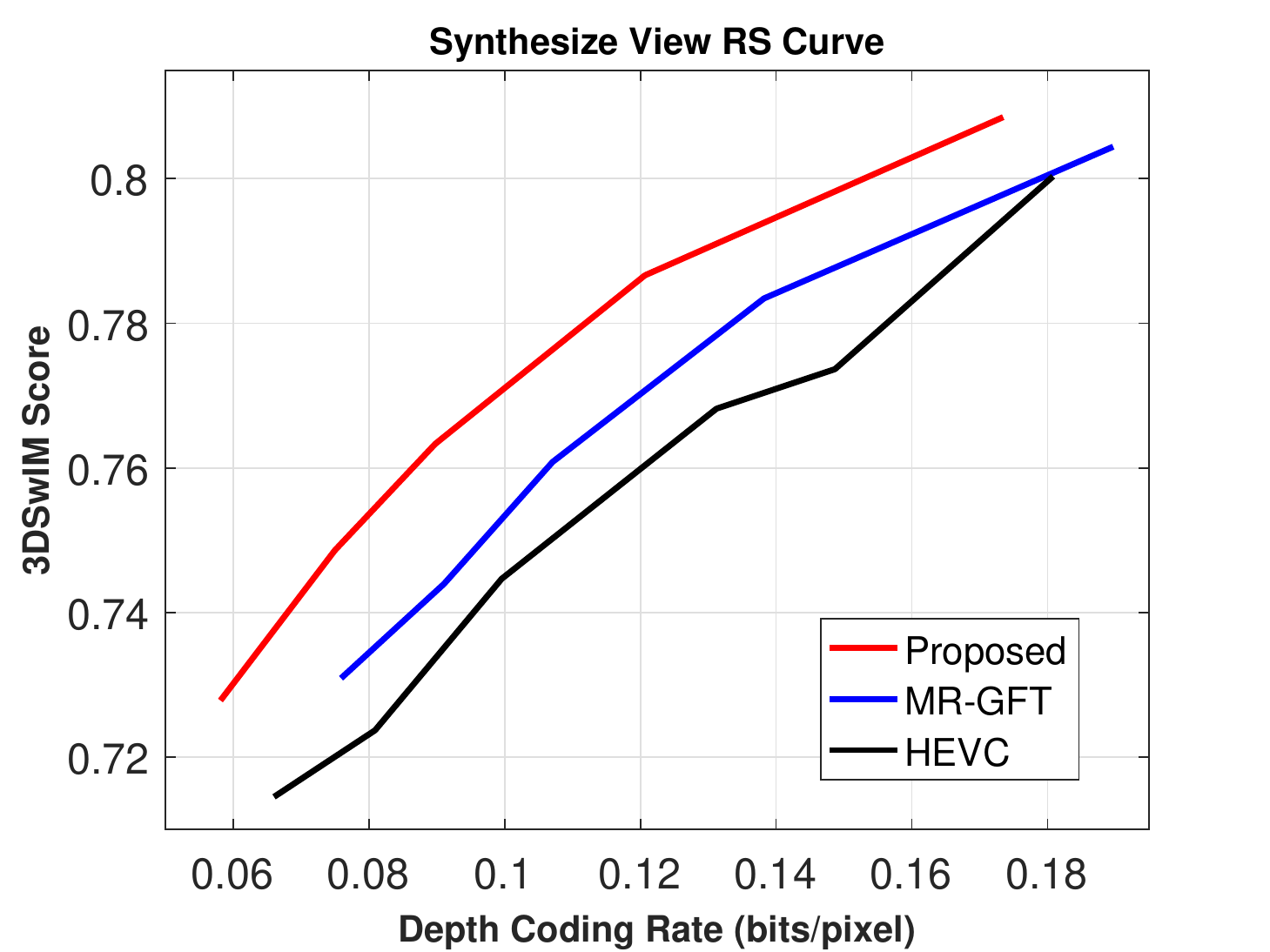}}
  \centerline{(c) \textrm{Lampshade}}\medskip
\end{minipage}
\hfill
\begin{minipage}[b]{.23\textwidth}
  \centering
  \centerline{\includegraphics[width=4.6cm]{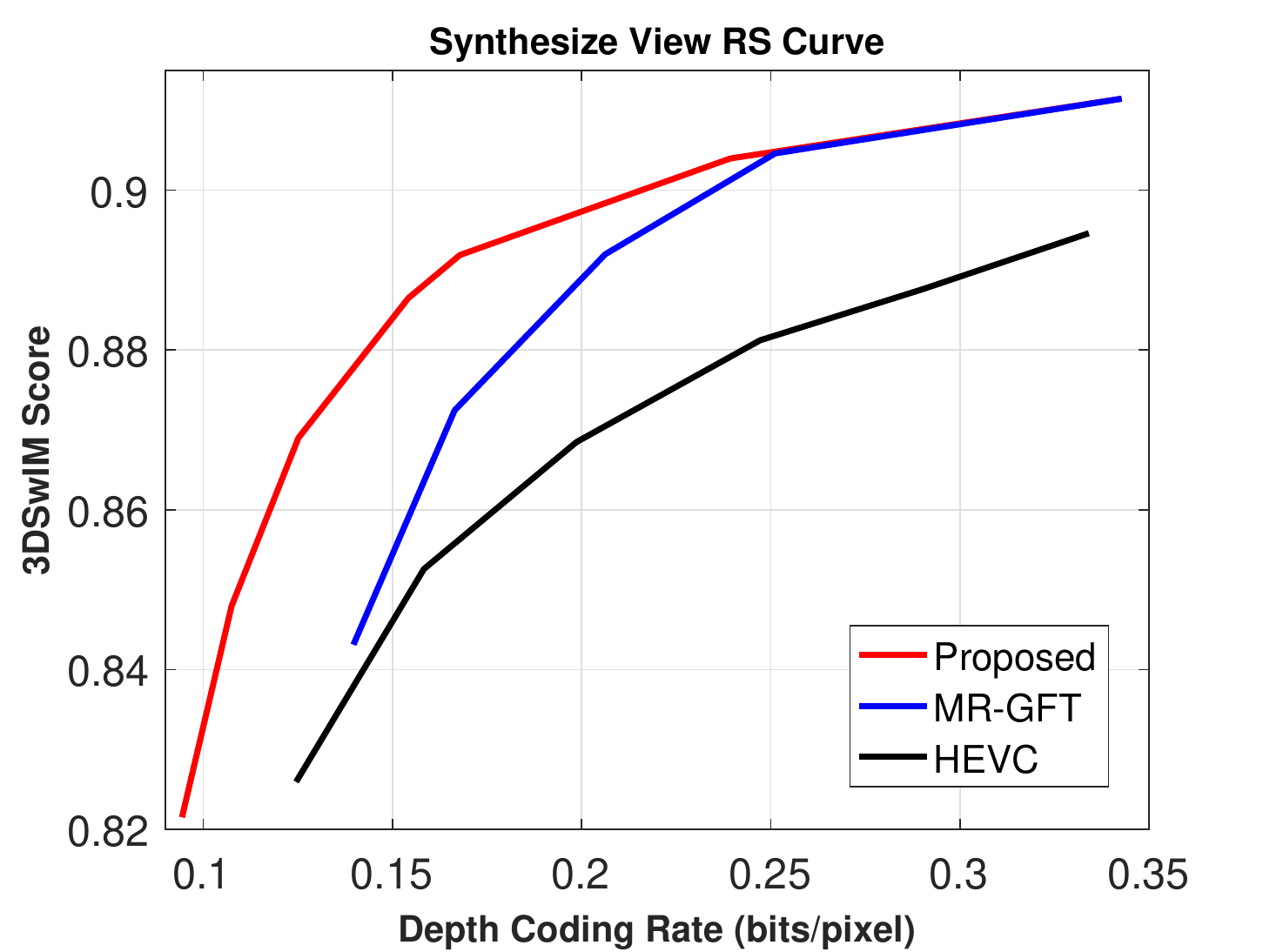}}
  \centerline{(d) \textrm{Aloe}}\medskip
\end{minipage}
\caption{Synthesized virtual view RS curves for \textrm{cones}, \textrm{Moebius}, \textrm{Lampshade} and \textrm{Aloe}, respectively.}
\label{fig:RS_Syn}
\end{figure*}

\begin{figure*}[t]
\begin{minipage}[b]{.23\textwidth}
  \centering
  \centerline{\includegraphics[width=4.6cm]{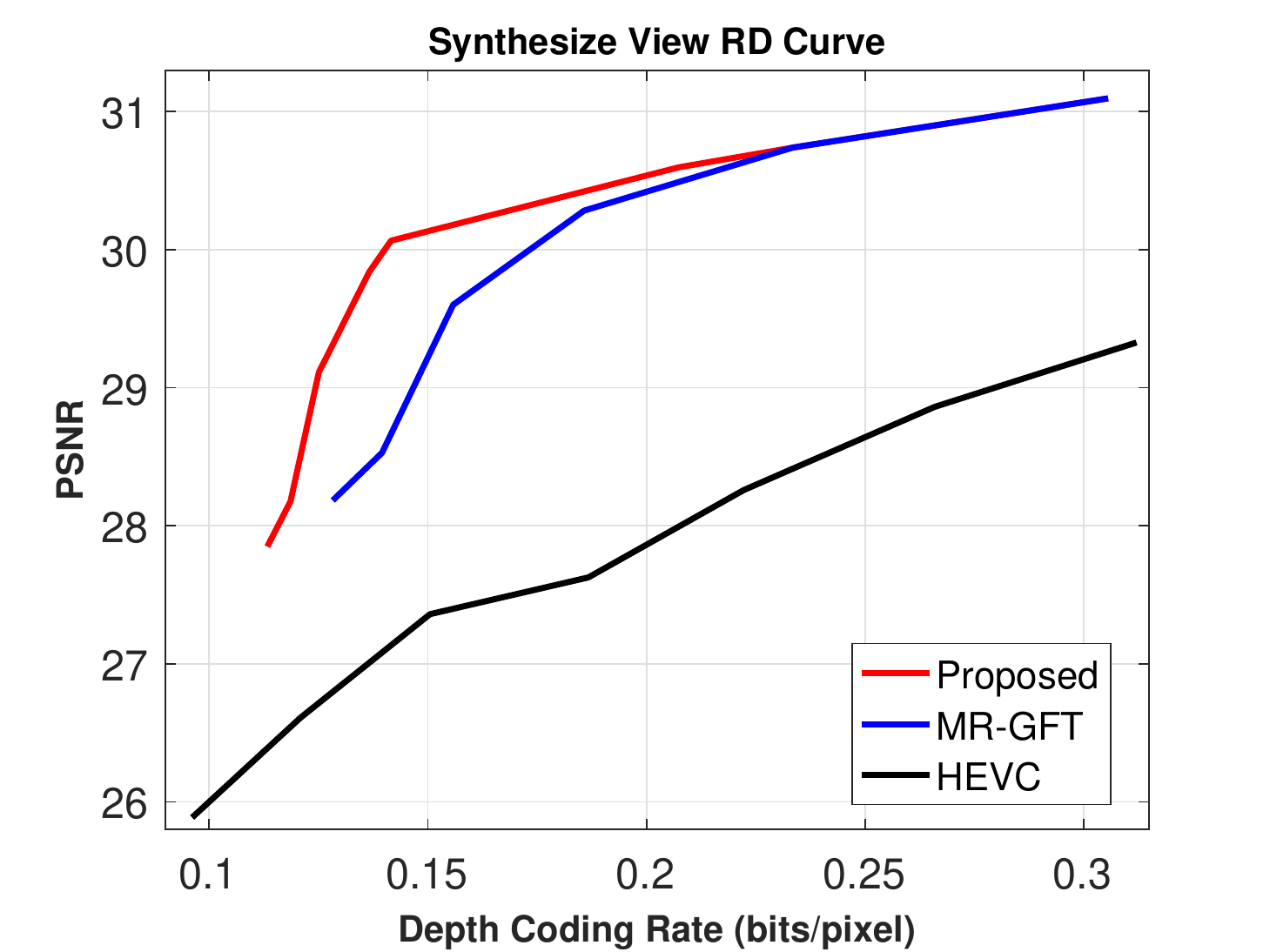}}
  \centerline{(a) \textrm{cones}}\medskip
\end{minipage}
\hfill
\begin{minipage}[b]{.23\textwidth}
  \centering
  \centerline{\includegraphics[width=4.6cm]{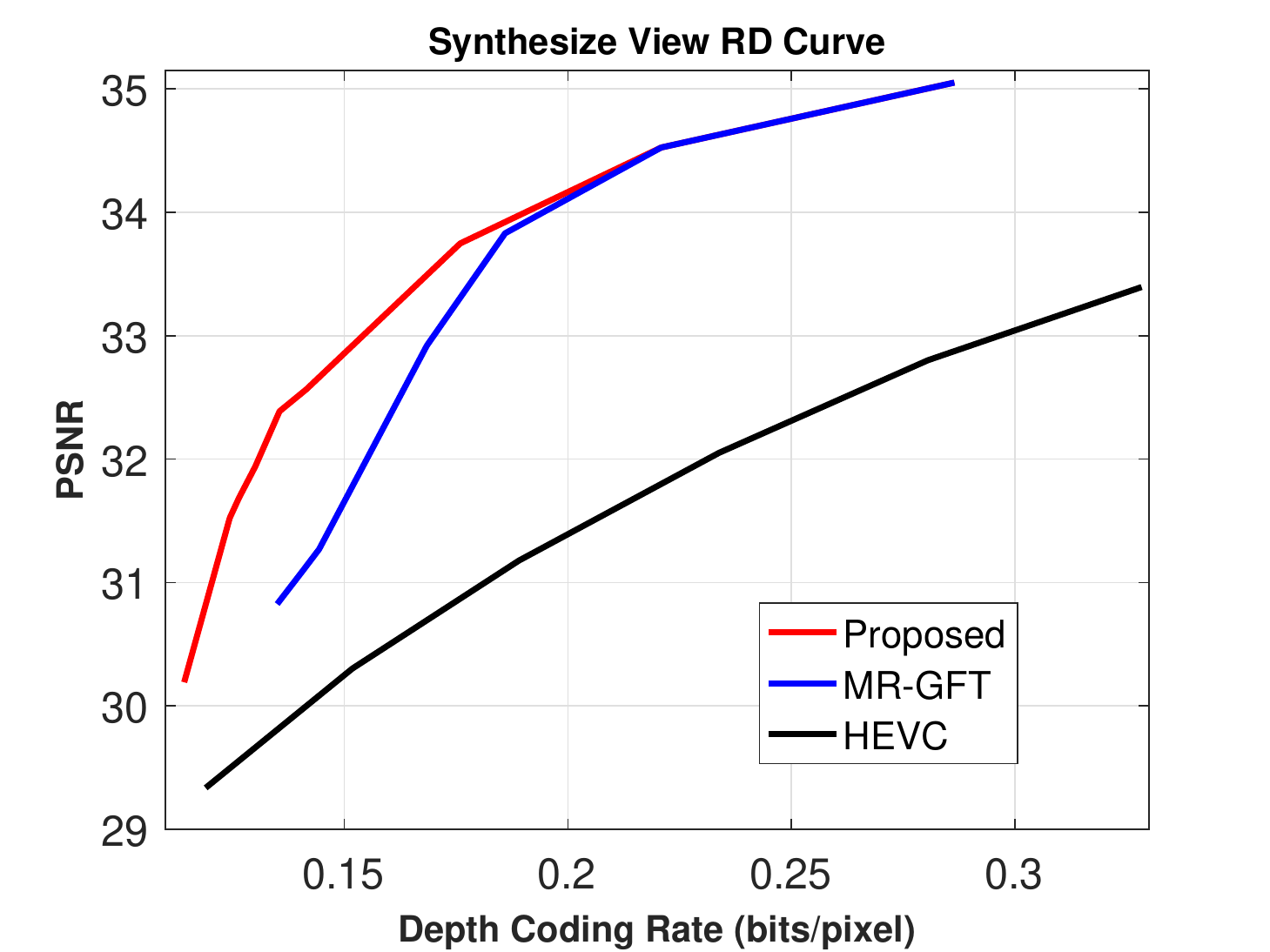}}
  \centerline{(b) \textrm{Moebius}}\medskip
\end{minipage}
\hfill
\begin{minipage}[b]{.23\textwidth}
  \centering
  \centerline{\includegraphics[width=4.6cm]{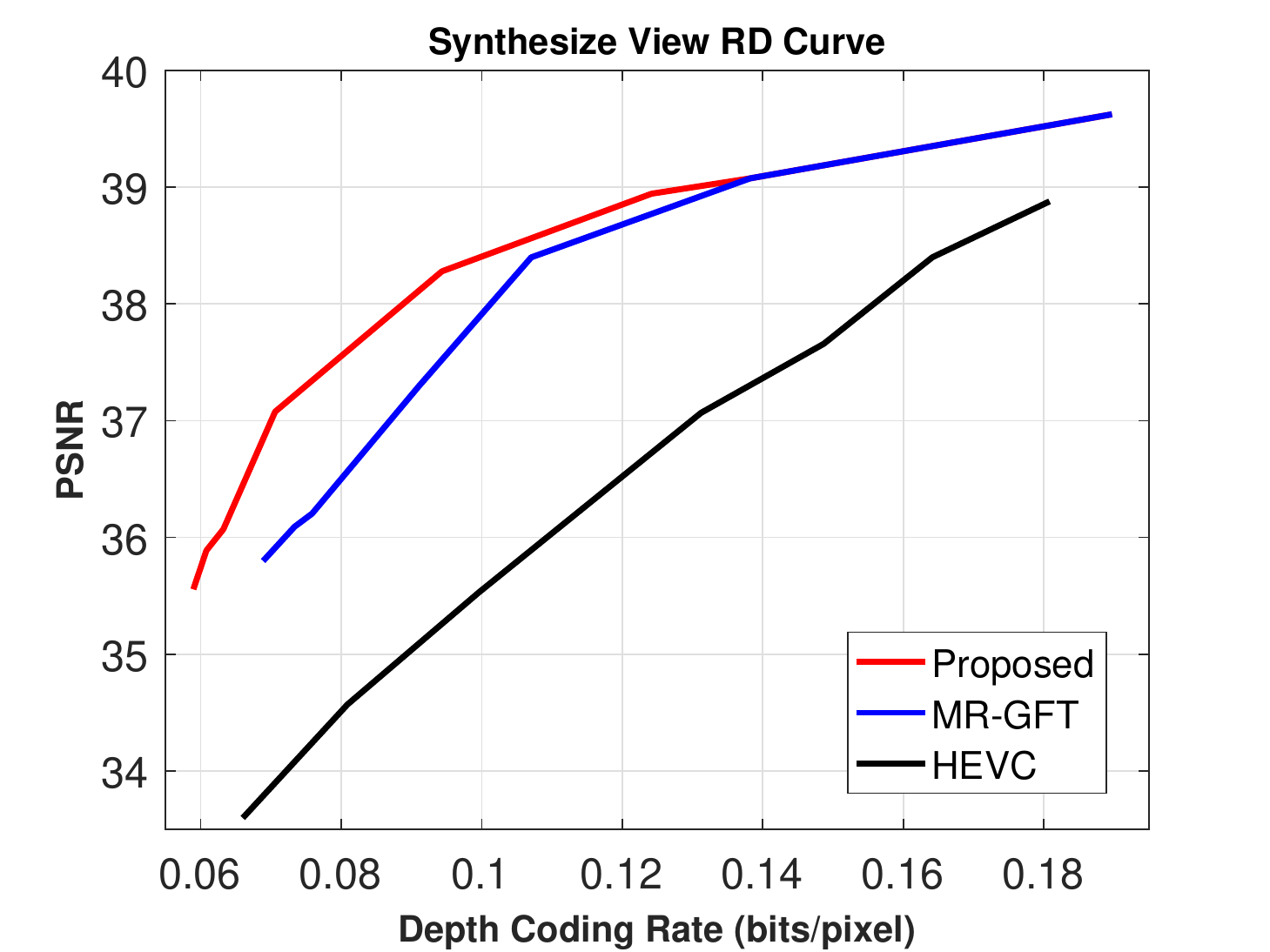}}
  \centerline{(c) \textrm{Lampshade}}\medskip
\end{minipage}
\hfill
\begin{minipage}[b]{.23\textwidth}
  \centering
  \centerline{\includegraphics[width=4.6cm]{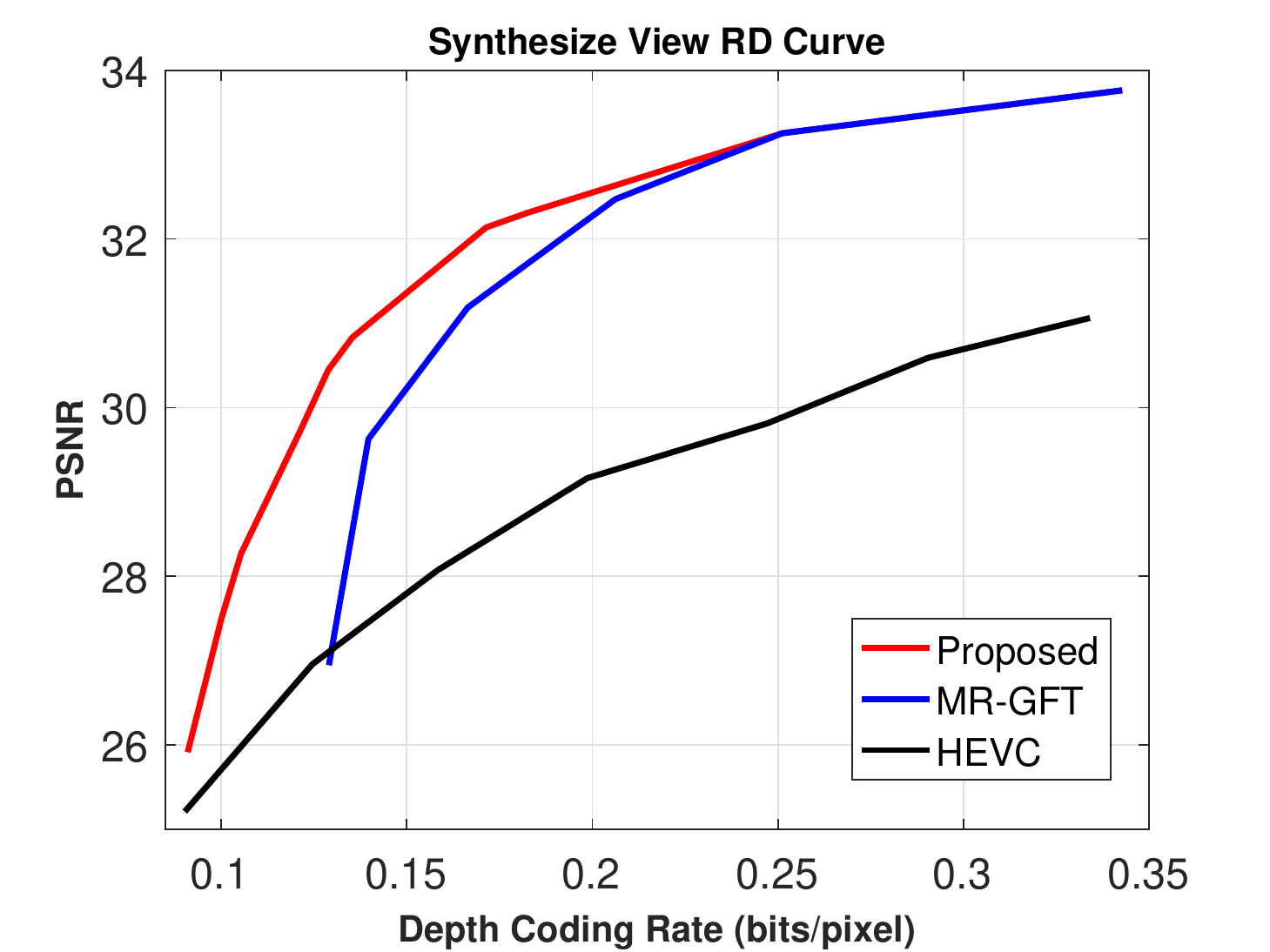}}
  \centerline{(d) \textrm{Aloe}}\medskip
\end{minipage}
\caption{Synthesized virtual view RD curves for \textrm{cones}, \textrm{Moebius}, \textrm{Lampshade} and \textrm{Aloe}, respectively.}
\label{fig:RD_Syn}
\end{figure*}

\begin{table*}
\centering
\caption{BG rate gain (RG) and PSNR gain (PG) for rate-score (RS) and rate-distortion (RD) curves}
\label{tab:Gain}
\begin{tabular}{|c|c|c|c|c|c|c|c|c|c|c|c|}
\hline
 \multicolumn{2}{|c|}{} & \textrm{teddy} & \textrm{cones} & \textrm{Dolls} & \textrm{Moebius}  & \textrm{Books}  & \textrm{Lamp.} &\textrm{Aloe} & \textrm{Dancer} & \textrm{GTFly} &\textbf{Avg.} \\
\hline

\multirow{2}{*}{RS-RG (rate\%)}
& \texttt{MR-GFT} & 15.78 & 26.10 & 19.38 & 19.01 & 29.69 & 18.35 & 18.82 & 7.31 & 8.52 & \textcolor{red}{18.11\%} \\
\cline{2-12}
& \texttt{HEVC} & 17.51 & 29.98 & 11.46 & 19.17 & 23.68 & 27.78 & 36.41 & 19.47 & 20.03 & \textcolor{red}{22.83\%} \\
\hline

\multirow{2}{*}{RD-RG (rate\%)}
& \texttt{MR-GFT} & 11.11 & 15.01 & 9.64 & 13.02 & 20.53 & 16.66 & 16.35 & 3.89 & 4.34 & \textcolor{red}{12.28\%} \\
\cline{2-12}
& \texttt{HEVC} & 39.54 & 52.90 & 27.70 & 40.60 & 43.68 & 44.51 & 38.55 & 45.30 & 48.29 & \textcolor{red}{42.34\%} \\
\hline

\multirow{2}{*}{RD-PG (dB)}
& \texttt{MR-GFT} & 0.55 & 0.92 & 0.19 & 1.03 & 0.47 & 1.29 & 1.20 & 0.23  & 1.55 & \textcolor{red}{0.83dB} \\
\cline{2-12}
& \texttt{HEVC} & 1.66 & 2.84 & 1.28 & 2.48 & 1.86 & 3.30 & 2.90 & 5.94 & 7.70 & \textcolor{red}{3.33dB} \\
\hline
\end{tabular}
\end{table*}

\subsection{Objective Results Compared to \texttt{MR-GFT} and \texttt{HEVC}}

To demonstrate the efficiency of our proposed contour approximation method (termed as \texttt{Proposed}), we compare our method with the \texttt{MR-GFT} method (depth images are compressed by multi-resolution GFT~\cite{hu2015tip} directly with the original detected contours that are losslessly coded).
We also consider using HEVC intra to compress the original depth images (termed as \texttt{HEVC}). 
The corresponding color images for both \texttt{MR-GFT} and \texttt{HEVC} methods are also compressed by HEVC intra. 
Since here we want to assess how object contour approximation affects depth image coding efficiency, we only consider the depth images with almost the same coding rate by all these three methods. For the corresponding color images, we select an identical $\text{QP}_\text{C}$ to compress them (meaning color images coding rates are also almost the same for these three methods). 
As recommend in the standard~\cite{muller20133d}, the QP offset $\Delta \text{QP}$ for the depth QP in relation to the color $\text{QP}_\text{C}$ is determined as $\text{QP} =\text{QP}_\text{C} + \Delta \text{QP}$, where $\Delta \text{QP} \leq 9$.
Finally, we synthesize virtual views using the decoded depth and color image pairs. 

Fig.\;\ref{fig:RS_Syn} shows the RS curves for \texttt{cones}, \texttt{Moebius}, \texttt{Lampshade} and \texttt{Aloe} image sets. 
The $x$-axis is the depth image coding rate in bits per pixel (bpp), and the $y$-axis is the synthesized view quality---scores measured by 3DSwIM. 
These figures show that our proposed method outperforms both \texttt{MR-GFT} and \texttt{HEVC}. 
The improved coding performance compared to \texttt{MR-GFT} validates the benefit of effectively approximating contours before edge-adaptive depth image compression.
It demonstrates that a depth contour can be properly approximated with little degradation to the synthesized view quality. 
Both our proposed method and \texttt{MR-GFT} have better performance than \texttt{HEVC}, since the multi-resolution GFT on PWS image compression is more efficient than DCT based transform~\cite{hu2015tip}.
Fig.\;\ref{fig:RD_Syn} illustrates the corresponding RD curves, where the $x$-axis is the depth coding rate and the $y$-axis is the PSNR value of synthesized virtual view. They also show that our proposed method has the best performance especially when the bit rate budget is low. 

\begin{figure*}[t]
\begin{minipage}{.23\textwidth}
  \centering
  \centerline{\includegraphics[bb=0 0 128 144, width=4.2cm]{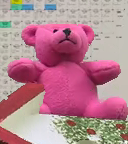}}
  \centerline{}
\end{minipage}
\hfill
\begin{minipage}{.23\textwidth}
  \centering
  \centerline{\includegraphics[bb=0 0 134 154, width=4.1cm]{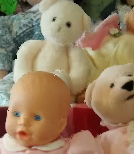}}
  \centerline{}
\end{minipage}
\hfill
\begin{minipage}{.23\textwidth}
  \centering
  \centerline{\includegraphics[bb=0 0 146 156, width=4.3cm]{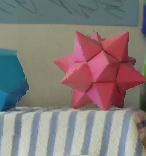}}
  \centerline{}
\end{minipage}
 \hfill
\begin{minipage}{0.23\textwidth}
  \centering
  \centerline{\includegraphics[bb=0 0 150 169, width=4.1cm]{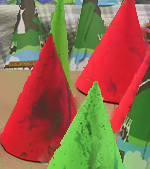}}
  \centerline{}
\end{minipage}
\hfill
\begin{minipage}{.23\textwidth}
  \centering
  \centerline{\includegraphics[bb=0 0 128 144, width=4.2cm]{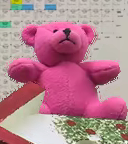}}
  \centerline{(a) \textrm{teddy}}
\end{minipage}
\hfill
\begin{minipage}{.23\textwidth}
  \centering
  \centerline{\includegraphics[bb=0 0 134 154, width=4.1cm]{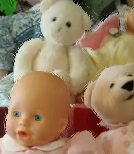}}
  \centerline{(b) \textrm{Dolls}}
\end{minipage}
 \hfill
\begin{minipage}{.23\textwidth}
  \centering
  \centerline{\includegraphics[bb=0 0 146 156, width=4.3cm]{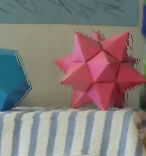}}
  \centerline{(c) \textrm{Moebius}}
\end{minipage}
\hfill
\begin{minipage}{.23\textwidth}
  \centering
  \centerline{\includegraphics[bb=0 0 150 169, width=4.1cm]{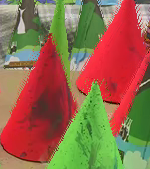}}
  \centerline{(d) \textrm{cones}}
\end{minipage}
\caption{Sub-regions of the synthesized views for \textrm{teddy}, \textrm{Dolls}, \textrm{Moebius} and \textrm{cones}, respectively. Images in the first row is synthesized by approximated and multi-resolution GFT compressed depth maps (proposed), where the second row is synthesized by HEVC compressed depth images. The depth and color coding rates are almost the same for each image sequence. (We strongly recommend readers to read an electronic version to distinguish the differences.)}
\label{fig:SynSameRate}
\end{figure*}

The computed BG gains~\cite{bjontegaard2001calcuation} are shown in Table~\ref{tab:Gain}. Comparing with \texttt{MR-GFT} and \texttt{HEVC}, we compute the rate gain (RG) (in percentage) for both RS and RD curves. We also compute the PSNR gain (PG) (in dB) for RD curve. 
The results in Table~\ref{tab:Gain} show that our proposed method can save a maximum of $29.69\%$ bit rate in RS measure compared to \texttt{MR-GFT} and the average rate gain achieves non-trivial $18.11\%$. We also achieve an average of $22.83\%$ rate gain in RS measure compared to \texttt{HEVC}. 
In RD measure, our method can also achieve $12.28\%$ rate gain and $0.83$dB PSNR gain compared to \texttt{MR-GFT}, and a $42.34\%$ rate gain and $3.33$dB PSNR gain compared to \texttt{HEVC}. These results prove the efficiency and effectiveness of our proposed method.

\subsection{Subjective Results Compared to \texttt{HEVC}}

We generated selected subjective results to visually examine images outputted by our proposed method, namely: sub-regions from the synthesized views of \texttt{teddy}, \texttt{Dolls}, \texttt{Moebius} and \texttt{cones}, as shown in Fig.~\ref{fig:SynSameRate}. 

In Fig.~\ref{fig:SynSameRate}, the first row corresponds to images synthesized from compressed color and depth images by our proposed method, and images on the second row are synthesized from compressed images by \texttt{HEVC}. 
In each column, the coding bit rates for the depth and color images by the two methods are almost the same. 
We observe that the synthesized images from our proposed method are more visually pleasing since edges in these images remain sharp. 
In contrast, there are noticeable bleeding effects around edges in the synthesized images from \texttt{HEVC}.
The results show that edge-adaptive depth image coding can lead to better synthesized view quality than fixed block transforms in compression standards like HEVC, which is consistent with results in previous edge-adaptive coding work \cite{hu2015tip,maitre08,sanchez09}.


\subsection{Results Compressed by 3D-HEVC intra}

The results demonstrated thus far show that our proposed object contour approximation can improve coding performance of edge-adaptive transform coding schemes like \cite{hu2015tip}. 
In theory, smoother contours can also improve other depth image codecs. 
To validate this point, we employ 3D video coding standard 3D-HEVC~\cite{muller20133d} intra HTM 6.0 to compress the original and the approximated color and depth images.

\begin{figure}[htb]
\begin{minipage}[b]{.23\textwidth}
  \hspace*{0.06in}
  \centering
  \centerline{\includegraphics[width=4.6cm]{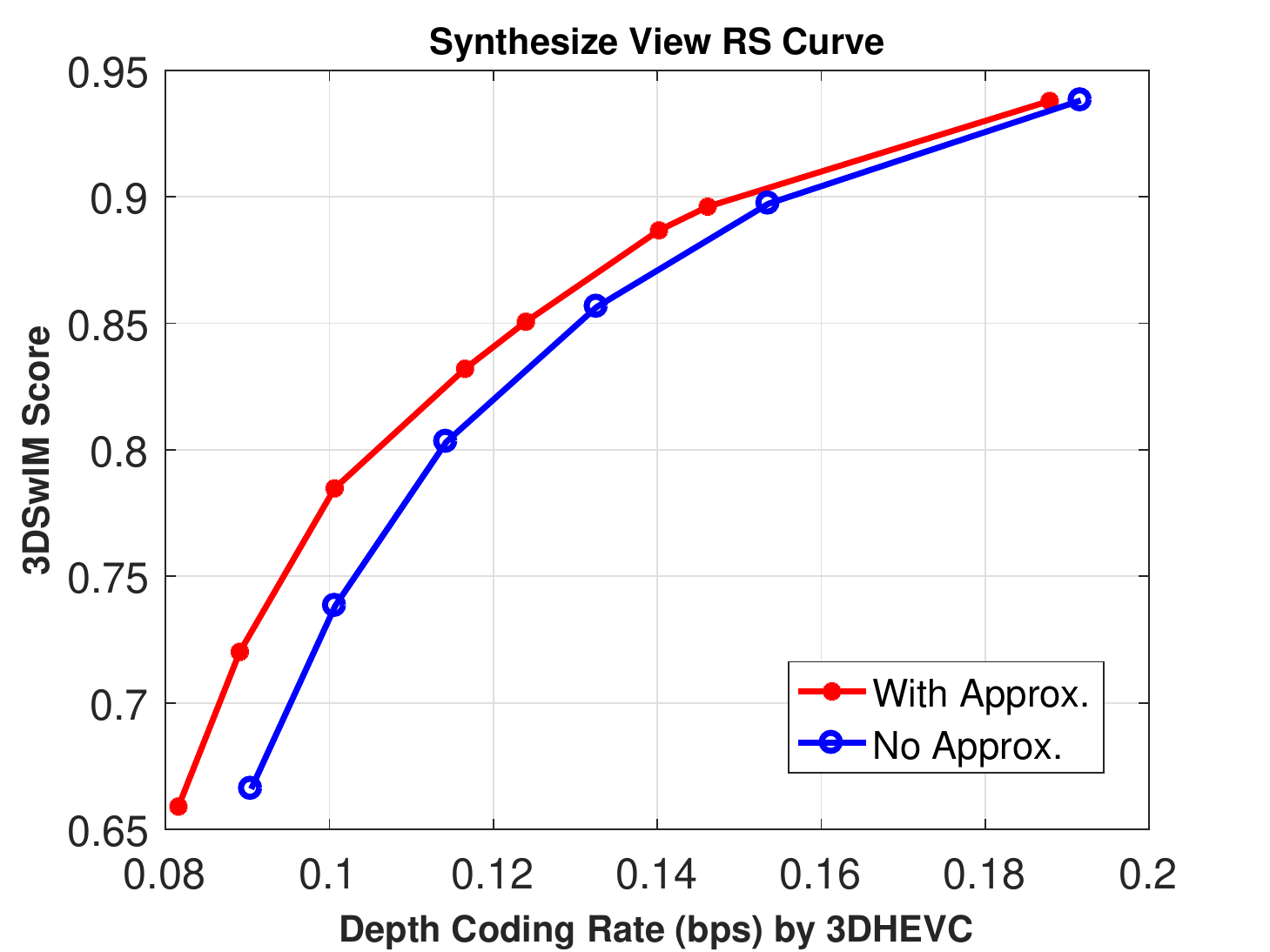}}
  \centerline{(d) \textrm{Undo Dancer}}\medskip
\end{minipage}
\hfill
\begin{minipage}[b]{.23\textwidth}
  \centering
  \centerline{\includegraphics[width=4.6cm]{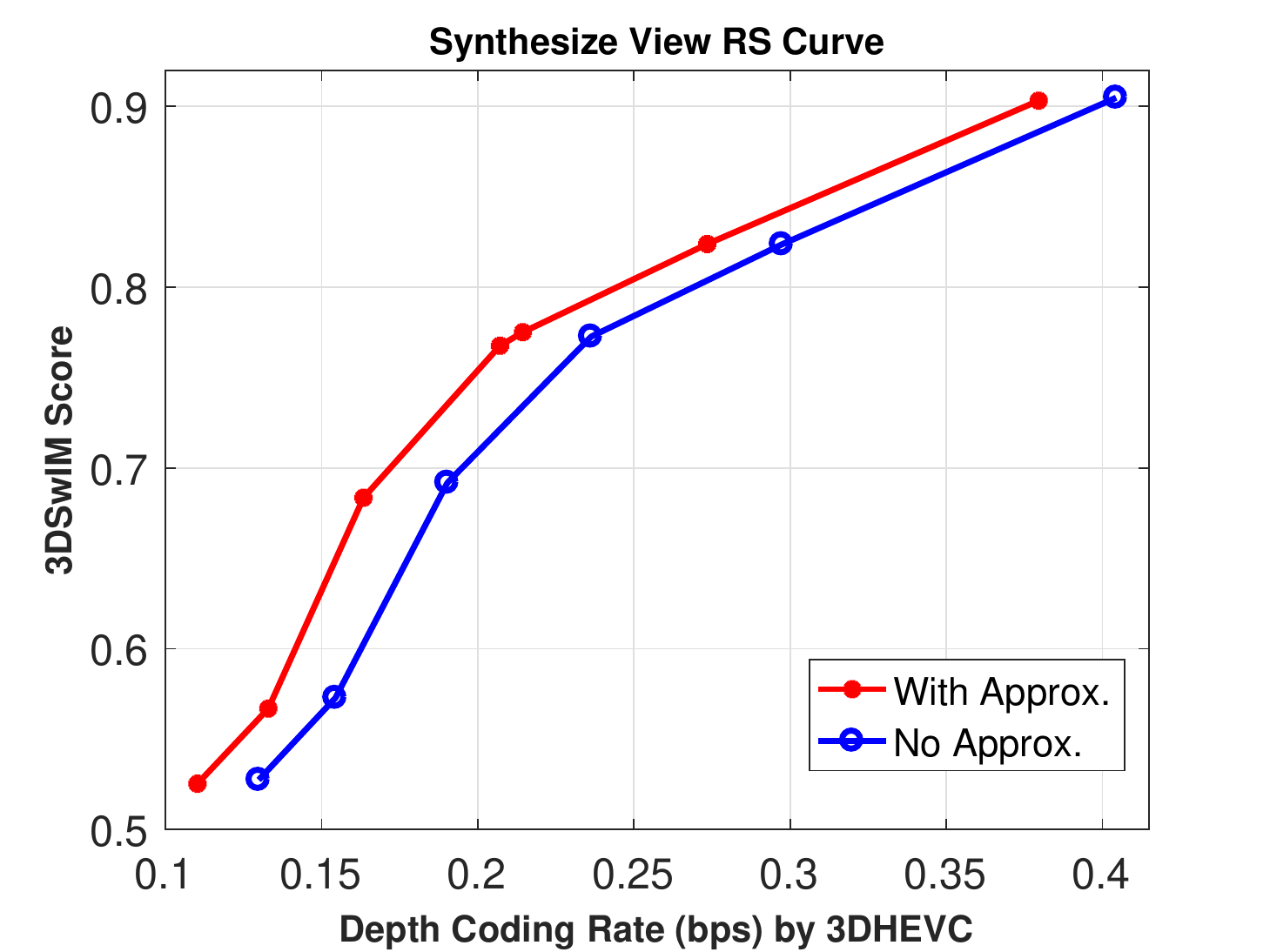}}
  \centerline{(d) \textrm{GTFly}}\medskip
\end{minipage}
\caption{RS curve by 3D-HEVC, where ``No Approx." and ``With Approx." mean using 3DHEVC to compress original and approximated color-plus-depth image pairs, respectively.}
\label{fig:3DHEVC}
\end{figure}

We test 3D-HEVC intra on \texttt{Undo Dancer} and \texttt{GTFly} sequences. 
We first approximate and alter the color-plus-depth image pairs with different values of $\lambda$. 
We then compress the original and the approximated color-plus-depth image pairs using 3D-HEVC, 
where the depth QP and color $\text{QP}_\text{C}$ pairs are the same for all the image pairs.
The resulting RS curves are shown in Fig.~\ref{fig:3DHEVC}. 
We see that using contour approximation, depth coding rate can be reduced by $6.84\%$ and $11.55\%$ for \texttt{Undo Dancer} and \texttt{GTFly} respectively. 
The coding gain can be explained as follows. 
In 3D-HEVC, a depth block is approximated by a model that partitions the block into two smooth sub-regions according to detected edges. 
The simplified (smoothed) contours can facilitate the partitioning of blocks into smooth sub-regions, resulting in lower depth coding rates.

\section{Conclusion}
\label{sec:conclude}
Efficient coding of depth image is essential for decoder-side virtual view synthesis via depth-image-based rendering (DIBR). Existing works employ either fixed transforms like DCT that blur a depth image's sharp edges at low rates, or edge-adaptive transforms that require lossless coding of detected edges as side information, which accounts for a large share of the bit budget at low rates. In this paper, we pro-actively alter object contours in an RD-optimal manner. We first propose a distortion proxy that is an upper bound of the established synthesized view quality metric, 3DSwIM. Given coding rate computed using arithmetic edge coding (AEC) and our distortion proxy, contours are approximated optimally via a dynamic programming (DP) algorithm in an inter-view consistent manner. With the approximated contours, depth and color images are subsequently augmented and coded using a multi-resolution codec based on GFT~\cite{hu2015tip} and HEVC respectively. 
Experiments show significant performance gain over previous coding schemes using either fixed transform or edge-adaptive transform with lossless coding of detected contours.

\appendix
\section{Appendix A}
\label{sec:appendix}

\subsection{Derivation of the maximum likelihood estimation of parameters $\sigma$ in (\ref{eq:pdfLaplace})}

Assuming the $M$ coefficients $\{c_1,\cdots,c_M\}$ are \textit{independent and identically distributed} and follow the probability density function (\ref{eq:pdfLaplace}), then the likelihood function for $M$ coefficients becomes
\begin{eqnarray}
L_\sigma(c) &= & \prod \limits_{i=1}^{M} \frac{1}{2\sigma} \mathrm{exp} \left(-\frac{|c_i|}{\sigma}\right) \\ \nonumber
&=& (2\sigma)^{-M} \cdot \mathrm{exp} \left( - \frac{1}{\sigma} \sum \limits_{i=1}^M |c_i| \right)
\label{eq:likelihood}
\end{eqnarray}
Take the $\mathrm{log}$ likelihood function as $l_\sigma(c) = \log (L_\sigma(c))$ and we get
\begin{eqnarray}
l_\sigma(c) = -M \ln(2\sigma) - \frac{1}{\sigma} \sum \limits_{i=1}^M |c_i|.
\end{eqnarray}
Take the derivative of $l_\sigma(c)$ with respect to $\sigma$
\begin{eqnarray}
\frac{\partial l}{\partial \sigma} = -\frac{M}{\sigma} + \frac{1}{\sigma^2}\sum \limits_{i=1}^M |c_i|.
\end{eqnarray}
To solve $\frac{\partial l}{\partial \sigma} = 0$, the maximum likelihood estimation of $\sigma$ is
\begin{eqnarray}
\sigma = \frac{1}{M} \sum \limits_{i=1}^M |c_i|. 
\end{eqnarray}
$\Box$

\bibliographystyle{IEEEtran}
\bibliography{ref}

\end{document}